\documentstyle[preprint,aps,amssymb,psfig,floats]{revtex}

\begin {document}

\draft

\title {Level Broadening Effects in Quantum Kinetic Equations:
        Hot Luminescence from a Quantum Wire near the Optical Phonon Emission
        Threshold}

\author {Michal Rokni}
\author {Y. Levinson}
\address {Department of Condensed Matter Physics, The Weizmann Institute of 
          Science, \\ Rehovot 76100, Israel}

\date {February 12, 1997}

\maketitle

\begin {abstract}
A recipe for the generalization of the Boltzmann equation to a quantum 
kinetic equation is given for cases in which only level shift and broadening
are considered, while coherence phenomena can be neglected. We also consider
a specific problem: Hot luminescence from a quantum wire near the threshold
for optical phonon emission. The problem is first discussed within the 
framework of the Boltzmann equation. After pointing out the failure of this 
description, the Boltzmann equations are generalized to a set of quantum 
kinetic equations, which in turn are solved in order to describe the 
luminescence spectra. 
\end {abstract} 

\pacs {}

\section {Introduction}
The simplest way to describe kinetic phenomena is by the Boltzmann equation 
(BE). There are however cases when the BE cannot be used, and one
has to use quantum kinetic equations (QKEs) that are in the simplest case 
equations for the one particle Green functions. If the Keldysh technique 
\cite {Kel65,LiP,RaS86} is used, the one particle Green functions that 
participate in the description of kinetic processes are the retarded Green 
function (and its complex conjugate), and the "statistical" Green function, 
which is simply the field correlator. For free particles these Green functions 
are diagonal in the basis of the free particle states, and are described by 
the free particle energies $ \epsilon_{\alpha} $, and their occupation numbers 
$ n_{\alpha} $. These are the quantities that appear in the BE for the 
{\em interacting} particles  {\em in external fields}.

The description of kinetic phenomena by QKEs instead of BEs results in two 
effects. The first effect is state mixing - due to interactions and external 
fields the Green functions are nondiagonal in the basis of the free particle 
states. This effect is responsible for coherence phenomena, of which a well 
known example is interband polarization described by semiconductor Bloch 
equations \cite{SCH88,LiK88,Kuz91a}.

The second effect is level shift and broadening - the Green functions are 
nonzero off the energy shell within some width defined by the interactions and 
the external fields. Level broadening effects in QKEs for phonons have 
been studied in Refs. \cite{Lev74,BKL76}.

The effects of level broadening are of importance in two cases. The first case
is that in which level broadening allows scattering processes that are 
forbidden by energy and momentum conservation. An example of such a case is 
that of electron-electron scattering in a 2DEG, in the presence of a quantizing
magnetic field \cite {Lev95}. If one tries to calculate the Auger transition 
rates responsible for Anti-Stokes luminescence, that was observed in Ref. 
\cite {PSM91}, using Fermi's golden rule, one encounters delta functions with 
an argument equal to zero, due to singularities in the density of states, see 
Ref. \cite {BRP93}. 

The second case in which level broadening effects should be considered is that 
of narrow energy distributions: When there are small energy scales in the 
problem that are narrower than the quantum width of the state. A simple 
example of such a case, that has one small energy scale, is that of an electron
distribution, excited by a narrow band-width laser. The problem becomes more 
complex when another small energy scale is added, for example: The electrons 
are excited to the close vicinity of some threshold. A problem of this nature
will be dealt with in this paper.

In the time domain, level shift and broadening lead to non-Markovian effects
in the scattering processes \cite{KaB,Lev70,Kuz91a}. These effects are 
important in the case of short excitation pulses 
\cite{Kuz91b,SBH94,BTR95,BKM96}.

If state mixing induced by coherent excitation can be neglected, and only shift
and broadening effects are considered, the QKEs can be presented in a form 
similar to the BEs. We will give a general recipe how to write such equations, 
and will use this recipe when considering the following problem: Hot 
luminescence from a quantum wire (QWR) due to electrons that are excited by a 
narrow band, noncoherent laser, to the close vicinity of the threshold for 
optical LO phonon emission.

Optical and transport phenomena in QWRs have been of great interest lately. 
The great interest in transport phenomena \cite {THS95,HTS95,YSW96,YSB97} is 
due to the possibility that a 1DEG in a QWR will have the properties of a 
Luttinger liquid. The Fermi edge singularity \cite {CGD91} and the possibility
of stimulated emission \cite {KHB89} enlarged interest in optics of QWRs . 
The important role played by LO phonons in the trapping of electrons 
\cite {WKC92} from three-dimensional extended states, into one-dimensional 
localized states, has been reported in Ref. \cite {RMK96}.

Theoretical studies of QWRs were concentrated around elementary excitations of 
a 1DEG \cite {HuD93,BMG96}, Fermi edge singularity \cite {RoM96},
excitonic effects \cite {OgT91a,OgT91b,BeH93}, electron-phonon scattering rates 
\cite {KnR93} and interactions \cite {SeD93a,SeD93b}, relaxation of 
photoexcited carriers \cite {RRL95}, calculation of envelope states 
\cite {SCG96}, and conductance of QWRs with self-consistent broadening effects
\cite {SWK90}.

\section {Quantum Kinetic Equations}
\label {QKE}

The retarded Green function and the "statistical" Green function which appear
in the kinetic equations are defined as
\begin {equation}
  \label {Gr-def}
  G^{\rm r}({\bf r}_{1}, \sigma_{1}, t_{1}; {\bf r}_{2}, \sigma_{2}, t_{2}) 
  \equiv 
  - i \theta (t_{1} - t_{2})   
  \langle 
    \left[ \hat {\Psi}({\bf r}_{1}, \sigma_{1}, t_{1}),
           \hat {\Psi}^{\dagger}({\bf r}_{2}, \sigma_{2}, t_{2}) \right]_{\pm}
  \rangle,
\end {equation}
\begin {equation}
  \label {Gs-def}
  G^{\rm s}({\bf r}_{1}, \sigma_{1}, t_{1}; {\bf r}_{2}, \sigma_{2}, t_{2}) 
  \equiv 
  - i \langle 
        \left[ \hat {\Psi}({\bf r}_{1}, \sigma_{1}, t_{1}),
               \hat {\Psi}^{\dagger}({\bf r}_{2}, \sigma_{2}, t_{2}) 
        \right]_{\mp}
      \rangle,
\end {equation}
where $ \hat {\Psi}({\bf r},\sigma, t) $ are the field operators in the 
Heisenberg representation. The square brackets with a plus (minus) sign signify
an anticommutator (commutator). The upper signs refer to Fermions and the 
lower signs refer to Bosons (this convention is used throughout the paper). 
The spin degrees of freedom $ \sigma $, will be suppressed from now on, and 
one can think of them as included in the space coordinates.

In case of interacting particles the Green functions obey the Dyson equation
\begin {eqnarray}
  \label {Dyson}
  \tilde {G}({\bf r}_{1}, t_{1}; {\bf r}_{2}, t_{2}) & = &
  \tilde {G}^{(0)}({\bf r}_{1}, t_{1}; {\bf r}_{2}, t_{2}) + 
  \nonumber \\ & &
  \int d {\bf r}_{3} \, d {\bf r}_{4} \int dt_{3} \, dt_{4} \,
  \tilde {G}^{(0)}({\bf r}_{1}, t_{1}; {\bf r}_{3}, t_{3})
  \tilde {\Sigma}({\bf r}_{3}, t_{3}; {\bf r}_{4}, t_{4})
  \tilde {G}({\bf r}_{4}, t_{4}; {\bf r}_{2}, t_{2}),
\end {eqnarray}
where $ \tilde {G} $ is a matrix of Green functions, and $ \tilde {\Sigma} $
is a matrix of self energy functions \cite {LiP}:
\begin {equation}
  \label {GStilde}
  \tilde {G} = \left(
                \begin {array}{cc}
                  0     & G^{\rm a} \\
                  G^{\rm r} & G^{\rm s} 
                \end {array}
               \right), \;\;\;\;\;
  \tilde {\Sigma} = \left(
                \begin {array}{cc}
                  \Sigma^{\rm s} & \Sigma^{\rm r} \\
                  \Sigma^{\rm a} & 0 
                \end {array}
               \right).
\end {equation}
The superscript $ (0) $ indicates free particle Green functions. The advanced 
Green function $ G^{\rm a} = G^{{\rm r} *} $.

In the case of free particles we can use the free particle eigenstates 
$ \psi_{\alpha}({\bf r}) $, and the corresponding annihilation operators 
$ \hat {a}_{\alpha}(t) $, to write 
$ \hat {\Psi}({\bf r}, t) = \sum_{\alpha} \psi_{\alpha}({\bf r}) \hat {a}_{\alpha}(t) $.
In this case $ G^{l(0)}({\bf r}_{1}, t_{1}; {\bf r}_{2}, t_{2}) $ depends only on the 
time difference $ t_{2} - t_{1} $, and can be Fourier transformed in time into
$ G^{l(0)}({\bf r}_{1}, {\bf r}_{2}; \epsilon) $ (the index $ l $ stands for 
r, a  or s). For an orthonormal set of states $ \alpha $ we can define 
$ G_{\alpha}^{l(0)}(\epsilon) $ such that
\begin {equation}
  \label{G-ae-def}
  G^{l(0)}({\bf r}_{1}, {\bf r}_{2}; \epsilon) = 
  \sum_{\alpha} \psi_{\alpha}({\bf r}_{1}) \psi_{\alpha}^{*}({\bf r}_{2})
  G_{\alpha}^{l(0)}(\epsilon).
\end {equation}

From the definitions (\ref {Gr-def}), (\ref {Gs-def}) we obtain (near the
singularities)
\begin {equation}
  \label {Gr0-ae}
  G^{{\rm r}(0)}_{\alpha}(\epsilon) = 
  \left[ 
    \begin {array}{c} 
      {\cal P}(\epsilon - \epsilon_{\alpha}) -
      i \pi \delta (\epsilon - \epsilon_{\alpha}) \\
      {\rm sign} (\epsilon)
      \left\{ 
       {\cal P}(\epsilon - {\rm sign} (\epsilon) \epsilon_{\alpha}) -
       i \pi \delta (\epsilon - {\rm sign} (\epsilon) \epsilon_{\alpha}) 
      \right\}
    \end {array} 
  \right],
\end {equation}
and
\begin {equation}
  \label {Gs0-ae}
  G^{{\rm s}(0)}_{\alpha}(\epsilon) = 
  \left[ 
    \begin {array}{c}
      - 2 \pi i ( 1 - 2 n_{\alpha} ) 
      \delta (\epsilon - \epsilon_{\alpha}) \\
      - 2 \pi i ( 1 + 2 n_{\alpha} )
      \delta (\epsilon - {\rm sign} (\epsilon) \epsilon_{\alpha})
    \end {array} 
  \right],
\end {equation}
where $ \epsilon_{\alpha} $ is the energy of state $ \alpha $, and 
$ {\cal P} $ is the principal part function. The upper (lower) term in the 
column corresponds to Fermions (Bosons). The occupation number of state 
$ \alpha $, 
$ n_{\alpha} = \langle \hat {a}_{\alpha}^{\dagger} \hat {a}_{\alpha}\rangle $,
is not necessarily the equilibrium occupation number.

In the case of interacting particles in external fields, if one neglects 
coherence effects, the Green functions are still diagonal in $ \alpha $, but the
energy levels are broadened and shifted. Let us first look at the time 
invariant case (systems under d.c.\ conditions). The level shift $ \Delta $, 
and the level broadening $ \Gamma $, are defined as
\begin {equation}
  \label {s-w-def}
  \Delta_{\alpha} (\epsilon) \equiv {\rm Re} 
  \Sigma^{\rm r}_{\alpha}(\epsilon), \;\;\;\;\;
  \Gamma_{\alpha}(\epsilon) \equiv - 2 \left[
     \begin {array}{c} 
       1 \\
       {\rm sign} (\epsilon)
     \end {array} \right]
     {\rm Im} \Sigma^{\rm r}_{\alpha}(\epsilon),
\end {equation}
where $ \Sigma^{\rm r}_{\alpha} $ is the retarded self energy of particles in 
state $ \alpha $. The column has the same meaning as in expression 
(\ref {Gr0-ae}).

The retarded Green function is an outcome of the Dyson equation (\ref {Dyson}),
that is written as
\begin {equation}
  \label {Dyson-Gr}
  G^{\rm r}_{\alpha}(\epsilon) = G^{{\rm r}(0)}_{\alpha}(\epsilon) +
  G^{{\rm r}(0)}_{\alpha}(\epsilon) \Sigma^{\rm r}_{\alpha}(\epsilon)
  G^{\rm r}_{\alpha}(\epsilon),
\end {equation} 
in the representation of the $ \alpha $ states.
Using the defintions (\ref {s-w-def}) we obtain from 
equation (\ref {Dyson-Gr}) (for positive frequencies)
\begin {equation}
  \label {Gr-ae}
  G^{\rm r}_{\alpha}(\epsilon) = 
  {\cal P} (\Gamma_{\alpha}(\epsilon) | \epsilon - \epsilon_{\alpha} -
                                          \Delta_{\alpha}(\epsilon)) -
      i \pi \delta (\Gamma_{\alpha}(\epsilon) | \epsilon - \epsilon_{\alpha} -
                                          \Delta_{\alpha}(\epsilon)),
\end {equation}
where
\begin {equation}
  \label {Pv}
  {\cal P} (\Gamma_{\alpha}(\epsilon) | \epsilon - \epsilon_{\alpha} -
                                          \Delta_{\alpha}(\epsilon)) =
  \frac { \epsilon - \epsilon_{\alpha} -  \Delta_{\alpha}(\epsilon) }
        { \left( \epsilon - \epsilon_{\alpha} - 
                 \Delta_{\alpha}(\epsilon) \right)^{2} +
          \left( \Gamma_{\alpha}(\epsilon) / 2 \right)^{2} },
\end {equation}
is the "smeared" and "shifted" principal value, and
\begin {equation}
  \label {delta}
  \delta (\Gamma_{\alpha}(\epsilon) | \epsilon - \epsilon_{\alpha} -
                                          \Delta_{\alpha}(\epsilon)) =
  \frac { \Gamma_{\alpha}(\epsilon) / 2 \pi }
        { \left( \epsilon - \epsilon_{\alpha} - 
                 \Delta_{\alpha}(\epsilon) \right)^{2} +
          \left( \Gamma_{\alpha}(\epsilon) / 2 \right)^{2} },
\end {equation}
is the "smeared" and "shifted" delta function. 

Comparing expression (\ref {Gr-ae}) with expression (\ref {Gr0-ae}) we see that 
$ G^{\rm r}_{\alpha}(\epsilon) $ is in fact a generalization of
$ G^{{\rm r}(0)}_{\alpha}(\epsilon) $, where the level shift and level width 
have been introduced into the principal value and delta functions. We therefore
write $ G^{\rm s}_{\alpha}(\epsilon) $ as a generalization of 
$ G^{{\rm s}(0)}_{\alpha}(\epsilon) $, thus defining the occupation functions
$ n_{\alpha}(\epsilon) $
\begin {equation}
  \label {Gs-ae}
  G^{\rm s}_{\alpha}(\epsilon) = 
  - 2 \pi i ( 1 \mp 2 n_{\alpha}(\epsilon) ) 
      \delta (\Gamma_{\alpha}(\epsilon) | \epsilon - \epsilon_{\alpha} -
                                      \Delta_{\alpha}(\epsilon)).
\end {equation}

Expressions (\ref {Gr-ae}) and (\ref {Gs-ae}) were written for the case of
positive frequencies, which will be assumed from now on. In case of negative 
Boson frequencies one has to multiply the retarded Green function and
$ \epsilon_{\alpha} + \Delta_{\alpha}(\epsilon) $ in these expressions by 
minus one.

The definitions above agree with the relation between the statistical Green 
function and the retarded Green function at thermal equilibrium
\begin {equation}
  \label {equil}
  G^{\rm s}_{\alpha}(\epsilon) = 2 i \left( 1 \mp 2 n_{T} (\epsilon) \right)
  {\rm Im} G^{\rm r}_{\alpha}(\epsilon),
\end {equation}
where $ n_{T} (\epsilon) $ is the equilibrium distribution function (Fermi or
Bose-Einstein).
The smeared delta function obeys the following normalization
\begin {equation}
  \label {norm}
  \int d \epsilon \,
  \delta (\Gamma_{\alpha}(\epsilon) | \epsilon - \epsilon_{\alpha} -
                                          \Delta_{\alpha}(\epsilon))
  = 1,
\end {equation}
and the occupation numbers are obtained from the occupation functions in the
following manner
\begin {equation}
  \label {on-of}
  n_{\alpha} = \int d \epsilon \,
  \delta (\Gamma_{\alpha}(\epsilon) | \epsilon - \epsilon_{\alpha} -
                                      \Delta_{\alpha}(\epsilon))
  n_{\alpha}(\epsilon).
\end {equation}

Relations (\ref {equil}) - (\ref {on-of}) can be obtained from the Lehmann 
representation of the Green functions. These relations are valid only if the 
one particle Green functions can be considered as diagonal in $ \alpha $, in 
other words, in the absence of coherence effects.

As a result of definitions (\ref {s-w-def}) and (\ref {Gs-ae}), the three 
unknown functions $ {\rm Re}G^{\rm r}, {\rm Im}G^{\rm r} $ and $ iG^{\rm s} $ 
have been replaced by a different set of three unknown functions: 
$ \Gamma_{\alpha}(\epsilon), \, \Delta_{\alpha}(\epsilon) $ and 
$ n_{\alpha}(\epsilon) $.

If the system is not under d.c.\ conditions, one can define a new pair of time
variables: the relative time $ t_{12} = t_{1} - t_{2} $, and the "center of 
mass" time $ \bar {t}_{12} = (t_{1} + t_{2})/2 $. Only slow processes will be
considered so that all quantities vary in $ \bar {t} $ with some time scale 
that is large compared to the inverse characteristic energy of the particles. 
In such a case one can still use the formalism given above for the Green 
functions, but with $ \bar {t} $ as a parameter of the problem, in addition to 
$ \epsilon $ (see for example Ref. \cite {LiP}).

For slow processes
\begin {equation}
  \label {Dyson-GrTe}
   G^{\rm r}_{\alpha}(\bar {t}, \epsilon) = 
   G^{{\rm r}(0)}_{\alpha}(\bar {t}, \epsilon) +
   G^{{\rm r}(0)}_{\alpha}(\bar {t}, \epsilon) 
   \Sigma^{\rm r}_{\alpha}(\bar {t}, \epsilon)
   G^{\rm r}_{\alpha}(\bar {t}, \epsilon).
\end {equation}
Again $ G^{\rm r}_{\alpha}(\bar {t}, \epsilon) $ can be written as a 
generalization of $ G^{{\rm r}(0)}_{\alpha}(\epsilon) $, in the same manner as 
before, but now $ \bar {t} $ is a parameter which appears in all functions. 
Thus, expressions (\ref {Gr-ae}) and (\ref {Gs-ae}) for 
$ G^{\rm r}_{\alpha} $ and $ G^{\rm s}_{\alpha} $, can be used with the three 
unknown functions 
$ \Gamma_{\alpha}(\bar {t}, \epsilon), \, 
  \Delta_{\alpha}(\bar {t}, \epsilon) $ and
$ n_{\alpha}(\bar {t}, \epsilon) $ that depend on $ \bar {t} $ as well as on
$ \epsilon $.

The known procedure for obtaining a kinetic equation for 
$ n_{\alpha}(\bar {t}, \epsilon) $ is to apply the operator 
$ \hat {G}^{-1*}_{\alpha2} - \hat {G}^{-1}_{\alpha 1} $ to the Dyson equation
for $ G^{\rm s}_{\alpha}(\bar {t}, t_{12}) $ \cite {RaS86}, where the operator 
$ \hat {G}^{-1}_{\alpha i} $ is equal to 
$ i \partial / \partial t_{i} - \epsilon_{\alpha} $ for electrons,
and $ \partial^{2} / \partial t_{i}^{2} + \epsilon_{\alpha}^{2} $ for photons
and phonons (we use the convention $ \hbar = 1 $ throughout the paper). The 
slow variation in $ \bar {t} $ should be considered and then the equation 
should be Fourier transformed in the time difference coordinates. This 
procedure leads to the QKE
\begin {equation}
  \label {qke}
  \frac {\partial}{\partial \bar {t}} G^{\rm s}_{\alpha}(\bar {t}, \epsilon) =
  - 2 \Sigma^{\rm s}_{\alpha}(\bar {t}, \epsilon)
  \, 
  {\rm Im} G^{\rm r}_{\alpha}(\bar {t}, \epsilon) +
  2 {\rm Im} \Sigma^{\rm r}_{\alpha}(\bar {t}, \epsilon) 
  \,
  G^{\rm s}_{\alpha}(\bar {t}, \epsilon).
\end {equation}
From now on the "bar" over $ t $ will be dropped, with the understanding that
all time variables are in fact center of mass time variables. Equation
(\ref {qke}) for photons and phonons includes a factor of 
$ \epsilon / \epsilon_{\alpha} $ that multiplies the time derivative on the
left hand side. Since we are interested in off-shell energies that are close to
the on-shell energy, this factor can be taken to be one.

In order to obtain equations for $ \Gamma_{\alpha}(t, \epsilon) $ and
$ \Delta_{\alpha}(t, \epsilon) $ one has to substitute an expression for
the retarded self energy function, in terms of the one particle Green 
functions, into the definitions of $ \Gamma $ and $ \Delta $ 
(\ref {s-w-def}). These equations are in fact the 
imaginary and the real parts of equation (\ref {Dyson-GrTe}). This yields 
coupled, self-consistent equations for $ \Gamma $ and $ \Delta $, that are
also $ n $ dependent. It is important to stress that the 
occupation function depends on time explicitly, while the dependence of 
$ \Gamma $ and $ \Delta $ on time is only through their dependence on $ n $. 
Therefore in order to find $ n $ one has to write a kinetic equation, that will
include time evolution through time derivatives. The equations for the level 
shift and width will include their dependence on time only through the 
appearance of $ n $, and will not include time derivatives.

If one uses the self-consistent-Born approximation for the self energy 
functions, one finds that the equations obtained for 
$ n_{\alpha}(t, \epsilon), \Gamma_{\alpha}(t, \epsilon) $, and
$ \Delta_{\alpha}(t, \epsilon) $ could have been obtained 
easily by applying a few generalizations to the BE for 
$ n_{\alpha} $. In the transfer from the BE to the QKE the number of unknowns 
increases from one ($ n $), to three ($ n, \Gamma $ and $ \Delta $), therefore
apart from the kinetic equation we require two more equations for the unknown 
functions $ \Gamma_{\alpha}(t, \epsilon) $ and 
$ \Delta_{\alpha}(t, \epsilon) $, that appear in the equation 
for $ n_{\alpha}(t, \epsilon) $.

We present here a recipe which allows one to go from the BE for 
$ n_{\alpha} (t) $, to the QKE for 
$ n_{\alpha}(t, \epsilon) $, and from the decay term in the Boltzmann 
equation (which is defined below) to an equation for 
$ \Gamma_{\alpha}(t, \epsilon) $. An equation for 
$ \Delta_{\alpha}(t, \epsilon) $ can be obtained from the
equation for $ \Gamma_{\alpha}(t, \epsilon) $ using Kramers-Kronig
relations.

The BE, in its most general form, can be written as
\begin {equation}
  \label {Boltzmann}
  \frac {\partial}{\partial t} n_{\alpha} =
  \sum_{\beta, \gamma ...} \left| M_{\alpha, \beta, \gamma ...} \right|^{2}
  \left[ (1 \mp n_{\alpha}) n_{\beta} (1 \mp n_{\gamma}) ... -
  n_{\alpha} (1 \mp n_{\beta}) n_{\gamma} ...
  \right] 
  2 \pi
  \delta (\epsilon_{\alpha} - \epsilon_{\beta} + \epsilon_{\gamma}...)\equiv S.
\end {equation}
This equation describes the evolution in time of the occupation number of 
particles in state $ \alpha $, due to interactions with particles in states 
$ \beta, \gamma ... $, with matrix elements $ M_{\alpha, \beta, \gamma ...} $.
Different states can also mean different particles. The right hand side of the 
equation includes delta functions that are responsible for energy conservation.

There are many cases in which it is convenient to think of the collision
integral $ S $ as composed of "scattering in" events and "scattering out" 
events: $ S = S_{\rm in} - S_{\rm out} $. The first term on the right hand 
side of equation (\ref {Boltzmann}) is the scattering in term: It describes 
processes in which particles enter state $ \alpha $ due to the interaction 
with other particles. This term can be written as 
$ S_{\rm in} = (1 \mp n_{\alpha}) {\cal G}_{\alpha} $, where 
$ {\cal G}_{\alpha} $ 
includes all the terms on the right hand side of the equation that do not 
multiply $ n_{\alpha} $. We refer to $ {\cal G}_{\alpha} $ as the "generation" 
term. The second term is the scattering out term: It describes processes in 
which particles leave state $ \alpha $. This term is usually written as 
$ S_{\rm out} = n_{\alpha}/ \tau_{\alpha} $, where $ \tau_{\alpha} $ is the 
scattering {\em out} time from state $ \alpha $. 

In the context of the generalization of the BE to a QKE, it is more convenient 
to think of the collision integral $ S $ as made up of "generation" and 
"decay" terms: 
$ S = {\cal G}_{\alpha} - \Gamma_{\alpha} n_{\alpha} $, where the decay term 
$ - \Gamma_{\alpha} n_{\alpha} $ includes all the terms containing 
$ n_{\alpha} $. Thus 
$ \Gamma_{\alpha} = \pm {\cal G}_{\alpha} + 1 / \tau_{\alpha} $ is 
the total decay rate of particles in state $ \alpha $. It is the {\em total} 
decay rate, and not only the scattering {\em out} rate, that is related to the 
retarded self energy \cite {KaB}. The general form for the decay rate 
$ \Gamma_{\alpha} $, as it is deduced from equation (\ref {Boltzmann}), is
\begin {equation}
  \label {decay}
  \Gamma_{\alpha} =
  \sum_{\beta, \gamma ...} \left| M_{\alpha, \beta, \gamma ...} \right|^{2}
  \left[ \pm n_{\beta} (1 \mp n_{\gamma}) ...
          +
  (1 \mp n_{\beta}) n_{\gamma} ...
  \right]
  2 \pi 
  \delta (\epsilon_{\alpha} - \epsilon_{\beta} + \epsilon_{\gamma}...).
\end {equation}
In a d.c.\ situation one obtains for the occupation numbers
\begin {equation}
  \label {n-a}
  n_{\alpha} = {\cal G}_{\alpha} / \Gamma_{\alpha}. 
\end {equation}

In order to transform the BE into a QKE
one should apply the following rules. To the left hand side of equation
(\ref {Boltzmann}) apply
\begin {itemize}
  \label {gen-lhs}
  \item
     $ n_{\alpha} (t) \rightarrow 
      \left( n_{\alpha}(t, \epsilon) \mp \frac {1}{2} \right)
      \delta (\Gamma_{\alpha}(t, \epsilon)| 
                              \epsilon - \epsilon_{\alpha} -
                              \Delta_{\alpha}(t, \epsilon)) $,
     where $ \epsilon $ is the off-shell energy of particles in state 
     $ \alpha $.
\end {itemize}
To the right hand side of the equation apply the following steps
\begin {itemize}
  \label {gen-rhs}
  \item 
     The occupation numbers for particles in states $ \alpha, \beta ... $ are
     replaced with the occupation functions, that depend on the
     off-shell energies $ \epsilon, \epsilon' ... $, corresponding to the 
     on-shell energies $ \epsilon_{\alpha}, \epsilon_{\beta} ... $ respectively:
     $ n_{\alpha} (t) \rightarrow n_{\alpha}(t, \epsilon) $,
     $ n_{\beta} (t) \rightarrow n_{\beta}(t, \epsilon') $...
  \item 
     The summand under the sum over all states $ \beta, \gamma ... $ is 
     multiplied by the product of the smeared delta functions for particles in 
     these states with the corresponding off-shell energies 
     $ \epsilon', \epsilon'' ... $, 
     and integrated over the off-shell energies: 
     \[ \sum_{\beta, \gamma ...} \rightarrow \sum_{\beta, \gamma ...}
       \int d \epsilon' \, d \epsilon'' ... 
       \delta (\Gamma_{\beta}(t, \epsilon')| 
                            \epsilon' - \epsilon_{\beta} -
                            \Delta_{\beta}(t, \epsilon'))
       \delta (\Gamma_{\gamma}(t, \epsilon'')| 
                            \epsilon'' - \epsilon_{\gamma} -
                            \Delta_{\gamma}(t, \epsilon'')) \times
     ... \]
  \item 
     The energy conservation delta function of the on-shell energies 
     $ \epsilon_{\alpha}, \epsilon_{\beta} ... $ is replaced with an energy
     conservation delta function of their corresponding off-shell energies 
     $ \epsilon, \epsilon' ... $:
     $ \delta (\epsilon_{\alpha} ...) \rightarrow \delta (\epsilon ...) $.
  \item 
     Multiply the entire right hand side by
     $ \delta (\Gamma_{\alpha}(t, \epsilon)| 
                              \epsilon - \epsilon_{\alpha} -
                              \Delta_{\alpha}(t, \epsilon)) $.
\end {itemize}

In order to obtain an equation for $ \Gamma_{\alpha}(t, \epsilon) $ one
should apply the first three steps of the generalization rules for the right 
hand side of the equation, to the expression for $ \Gamma_{\alpha} $ 
(\ref {decay}).

The QKE can be written generally in the form
\begin {eqnarray}
  \label {qke-GG}
  & & 
  \frac { \partial }{ \partial t }
  \left\{
     \left( n_{\alpha}(t, \epsilon) \mp \frac {1}{2} \right)
     \delta (\Gamma_{\alpha}(t, \epsilon)| \epsilon - \epsilon_{\alpha} -
                                \Delta_{\alpha}(t, \epsilon))
  \right\} = \nonumber \\ 
  & &
  \left\{ {\cal G}_{\alpha}(t, \epsilon) - 
  \Gamma_{\alpha}(t, \epsilon) n_{\alpha}(t, \epsilon)
  \right\}
  \delta (\Gamma_{\alpha}(t, \epsilon)| \epsilon - \epsilon_{\alpha} -
                                \Delta_{\alpha}(t, \epsilon)).
\end {eqnarray}
The function $ {\cal G}_{\alpha}(t, \epsilon) $ contains all the terms 
on the right hand side of the QKE, that do not multiply 
$ n_{\alpha}(t, \epsilon) $. It is the generating term - the term in the
kinetic equation that is responsible for the particle generation rate. Equation
(\ref {qke-GG}) shows clearly that the particle level width 
$ \Gamma_{\alpha}(t, \epsilon) $ is in fact the total particle decay 
rate. Note that due to (\ref {norm}) the integration of equation 
(\ref {qke-GG}) over $ \epsilon $ results in the disappearance of the 
$ \mp 1/2 $ factor from the left hand side. The occupation function in the
time independent case is given by
\begin {equation}
  \label {n-ae}
  n_{\alpha} (\epsilon) = 
  {\cal G}_{\alpha} (\epsilon) / \Gamma_{\alpha} (\epsilon).
\end {equation}

The representation of the QKE in terms of $ n, \Gamma $
and $ \Delta $ is more convenient than the representation in terms
of Green functions, since one can clearly see all physical processes, and use
the physical intuition that one gained from the BE in order to 
simplify the QKE.

\section {Luminescence from a Quantum Wire}
\label {Lum}
\begin {figure}
  \begin{center}
    \setlength{\unitlength}{0.0925in}
    \begin{picture}(36,32)(0,0)
      \put(0,0){\psfig {figure=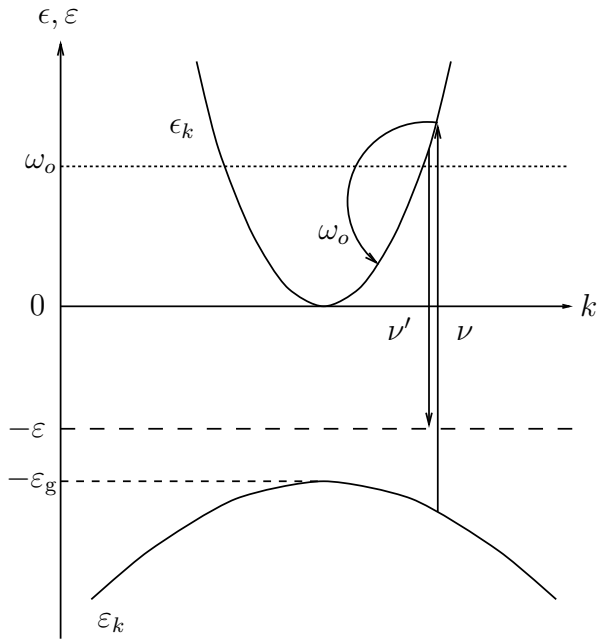,width=30\unitlength}}
      \put(3,1){\makebox(0,0){$\varepsilon_{k}$}}
      \put(7,29){\makebox(0,0){$\epsilon_{k}$}}
      \put(-1,27){\makebox(0,0){$\omega_{o}$}}
      \put(-1.2,19){\makebox(0,0){$0$}}
      \put(0,35){\makebox(0,0)[b]{$\epsilon,\varepsilon$}}
      \put(-1.8,12){\makebox(0,0){$-\varepsilon$}}
      \put(-1.5,9){\makebox(0,0){$-\varepsilon_{\rm {g}}$}}
      \put(16.5,23){\makebox(0,0)[r]{$\omega_{o}$}}
      \put(20,17){\makebox(0,0)[br]{$\nu'$}}
      \put(22.5,17){\makebox(0,0)[bl]{$\nu$}}
      \put(30,19){\makebox(0,0){$k$}}
   \end{picture}
  \end{center}
  \caption {Hot luminescence from a QWR: Electrons are excited from
            the highest valence subband $ \varepsilon_{k} $, to the lowest 
            conduction subband $ \epsilon_{k} $, via photons $\nu$. These 
            electrons relax to the bottom of the band emitting LO phonons 
            $\omega_{o}$. The hot luminescence $\nu'$ is due to recombination 
            of electrons from the vicinity of the threshold for phonon emission
            (dotted line) with holes in an impurity level $ -\varepsilon $ 
            (large dashed line).}
  \label {qwr}
\end {figure}
We now use the formalism given in section \ref {QKE} in order to deal with the 
specific problem of hot luminescence from a QWR. The case in which electrons 
are excited from the highest valence subband to the lowest conduction subband 
via photons, and then relax via LO phonon emission (see Fig. \ref {qwr}) is
considered. All other conduction and valence subbands are neglected (in 
contrast to Ref. \cite {RRL95,RoM96}). We consider a situation in which 
electrons are excited by a narrow band, noncoherent laser, just above the 
threshold for optical phonon emission $ \epsilon_{k} = \omega_{o} $, and 
describe the hot luminescence due to the recombination of electrons with holes 
in an impurity level. It is assumed there is some mechanism due to which the 
electrons leave the bottom of the conduction subband, so that there is no 
electron accumulation there, but this recombination can be neglected in the 
balance equations near the threshold. The spin degree of freedom will be 
completely disregarded.

Unless specified otherwise we will assume that the electrons are excited
above the threshold, since we are mainly interested in luminescence due to 
these electrons. Below the threshold the electrons relax to the bottom
of the band by emitting acoustical phonons. The relaxation rate due to
acoustical phonons $ \tau_{\rm ac}^{-1} $, is much smaller than that due to 
optical phonons, and can be neglected above the threshold.

The QWR runs along the $ z $ direction so that the electron wave functions are 
localized in the $ xy $ plane, and free waves in the $ z $ direction. The wave 
function in the conduction band can be written as 
$ e^{i k z} \phi^{\rm e} ({\bf r}) / L^{1/2} $, where $ L $ is the wire 
normalization length, and $ {\bf r} $ is a vector in the $ xy $ plane.
The electron energy is $ \epsilon_{k} = k^{2}/ 2 m_{\rm e} $. All energies 
are measured from the bottom of the conduction band. The holes in the valence 
band have wave functions of the form 
$ e^{i p z} \phi^{\rm h} ({\bf r}) / L^{1/2} $ and energy
$ \varepsilon_{\rm g} + \varepsilon_{p} $, with 
$ \varepsilon_{p} = p^{2}/ 2 m_{\rm h} $.

The exciting photons are taken to be plane waves and the photon frequency is 
$ \nu = \nu_{{{\bf f}}} = c |\bf f| $ ($ {{\bf f}} $ is the photon momentum 
and $ c $ is the light velocity). The LO phonons are three-dimensional and 
have a flat dispersion law, $ \omega_{{\bf q}} = \omega_{o} $.

The impurities are distributed randomly in the bulk. The wave function of 
a hole in impurity $ i $ is of the form 
$ \psi ({\bf r} - {\bf r}_{i}, z - z_{i}) $, 
where $ {\bf r}_{i}, z_{i} $ are the coordinates of the impurity position. 

In order to analyze the hot luminescence we neglect light polarization effects 
for simplicity, and assume that the wire and the crystal are cylindrically 
symmetric (an assumption that would not fit V-grooved wires nor the assumption 
of Ref. \cite {SCG96}). Thus, the wave functions of the emitted photons are 
\begin {equation}
  \label {ph-we}
  \chi_{f,m,n} ({\bf r},z) =
  \left( \frac {1}{\pi R^{2} L \left| J_{m+1}(\kappa_{m,n} R) \right|^{2}}
  \right)^{1/2} 
  e^{i f z} e^{i m \varphi} J_{m}(\kappa_{m,n} r),
\end {equation} 
where $ R $ is the normalizing radius and $ L $ the normalizing length of the
crystal, and $ z, \varphi $ and $ r $ are the cylindrical polar coordinates.
$ J_{m} $ are the Bessel functions, and $ (\kappa_{m,n} R) $ is the 
$ n^{\rm th} $ 
zero of $ J_{m} $. The luminescence photon frequency is given by 
$ \nu' = \nu_{f,m,n} = c \left( f^{2} + \kappa_{m,n}^{2} \right)^{1/2} $
(where the prime signifies this frequency is of luminescence, and not of 
excitation). We neglect all excitonic effects (these are treated for the case 
of optical absorption in one-dimensional semiconductors in Refs. 
\cite {OgT91a,OgT91b,BeH93}).

We begin by writing down the BEs for the electrons in the conduction band, and 
for the luminescence photons. It is shown why this description fails and one 
has to turn to the QKEs. We will then employ the generalization rules in order 
to obtain the QKEs from the BEs, and use the QKEs in the analysis of the 
luminescence.

\subsection {The Boltzmann Description}
\label {B-Des}
We assume low temperature and a weak excitation level. At equilibrium, when
there is no excitation, all the electrons are in the valence band, and there
are no phonons ($ N(\omega_{o}) = 0 $), due to the low temperature. In other
words, there are no electrons in the conduction band ($ n(k) = 0 $), no 
holes in the valence band ($ \bar {n}(p) = 0 $), and the impurity level
is fully occupied by holes ($ \bar {n}(\varepsilon) = 1 $).

The excitation creates electrons in the conduction band, and holes in the
valence band. Most of the excited electrons relax to the bottom of the 
conduction band, emitting LO phonons, but a small fraction of them recombine 
with holes in the impurity level, producing hot luminescence. By the assumption 
of weak excitation we mean that $ N(\omega_{o}) $, $ n(k) $ and $ \bar {n}(p) $
are small. The luminescence would be even weaker than the excitation, therefore 
$ \bar {n}(\varepsilon) $ remains close to one. Due to the weak excitation 
the luminescence is spontaneous, and the occupation numbers for the 
luminescence photons are also small.

In the derivation of the BEs, terms that are nonlinear in the electron 
occupation numbers will be neglected. We will also neglect nonequilibrium 
contributions to the phonon and the hole (both in the valence band and in the 
impurity level) occupation numbers.

The interaction of the electrons with the luminescence field is neglected in 
the balance equation for the electrons, in comparison with their interaction 
with the excitation field, due to the weakness of the excitation. In the case 
of c.w.\ excitation the balance equation for the electrons is
\begin {eqnarray}
  \label {e-Be}
  0 = \sum_{{\bf f}} \sum_{p}
  \left|M^{\rm exc}_{k,p,{\bf f}}\right|^{2}
  2 \pi 
  \delta (\nu_{{\bf f}} - \epsilon_{k} - \varepsilon_{\rm g} - \varepsilon_{p})
  (1 - n(k)) N({\bf f}) - \nonumber \\
  \sum_{{\bf q}} \sum_{k'}
  \left|M^{\rm e-LO}_{k,k',{\bf q}}\right|^{2}
  2 \pi \delta (\epsilon_{k} - \epsilon_{k'} - \omega_{o})
  n (k),
\end {eqnarray}
where $ M^{\rm exc}$ is the matrix element for the 
electron-exciting-photon interaction, and $ M^{\rm e-LO} $ is the 
matrix element of the electron-LO phonon interaction.

Let us first look at the electron-exciting-photon matrix element. In this
interaction one photon of wave vector $ {\bf f} $ is absorbed, and a hole of 
momentum $ p $ and an electron of momentum $ k $ are created. The interaction, 
in second quantized form, is given by 
\begin {equation}
  \label {int-exc}
  \int d {\bf r} \int dz \, \frac { e p_{o} }{ m_{o} c }
  \hat \Psi^{{\rm e} \dagger}({\bf r},z) 
  \hat {A}^{\rm exc}({\bf r}, z) \hat \Psi^{{\rm h} \dagger}({\bf r},z).
\end {equation}
The mass $ m_{o} $ is the bare electron mass and the constant $ p_{o} $ is the 
"bare" electron momentum operator (in the direction of the photon polarization,
assumed linear for simplicity), sandwiched between the valence band top and the
conduction band bottom Bloch wave functions. $ \hat \Psi^{\rm e}({\bf r},z) $
and $ \hat \Psi^{\rm h}({\bf r},z) $ are the electron and hole field operators,
respectively. The photon field operator is given by 
\begin {equation}
  \label {exc-fld}
  \hat {A}^{\rm exc}({\bf r}, z) =
  \sum_{{\bf f}}
  \left( \frac {2 \pi c^{2} }{ V \nu_{{\bf f}} } \right)^{1/2}
  e^{i {\bf f}_{\perp} \cdot {\bf r} + i f z} 
  \left( \hat {a}_{{\bf f}} + \hat {a}_{-{\bf f}}^{\dagger} \right),
\end {equation}
where $ {\bf f}_{\perp} $, $ f $ are the components of $ {\bf f} $ in the $ xy $ plane, and in the $ z $ direction, respectively, $ \hat {a}_{{\bf f}} $ is the 
annihilation operator, and $ V $ is the normalization volume of the crystal in 
which the quantum wire is embedded.

In order to find $ M^{\rm exc} $ one should sandwich interaction 
(\ref {int-exc}) between the states
 $ \langle n(k)=1, \bar{n}(\varepsilon_{p})=1, N_{{\bf f}}= 0 | $ 
and
 $ | \left. n(k)=0, \bar{n}(\varepsilon_{p})=0, N_{{\bf f}}= 1 \rangle \right. $. 
This is equal to
\begin {equation}
  \label {M-e-ep}
  M^{\rm exc}_{k,p,{\bf f}} = \frac {e p_{o}}{m_{o} c}
  \left( \frac {2 \pi c^{2}}{V \nu_{{\bf f}}} \right)^{1/2} 
  \frac {1}{L} \int d {\bf r} \int dz \,
  e^{-i (k + p) z} \phi^{{\rm e} *} ({\bf r}) \phi^{{\rm h} *} ({\bf r})
  e^{i {\bf f}_{\perp} \cdot {\bf r} + i f z}.
\end {equation}
The photon momentum is small and so is $ {\bf r} $, since it is limited by the
wire cross section, therefore $ e^{i {\bf f}_{\perp} \cdot {\bf r}} = 1 $. 
Integrating over the $ z $ coordinates the square of the matrix element is
\begin {equation}
  \label {|M-e-ep|^2}
  \left| M^{\rm exc}_{k,p,{\bf f}} \right|^{2} = 
  \left( \frac {e p_{o}}{m_{o} c} \right)^{2}
  \frac {2 \pi c^{2}}{V \nu_{{\bf f}}}
  \left| \int d {\bf r} \, 
         \phi^{\rm e} ({\bf r}) \phi^{\rm h} ({\bf r})
  \right|^{2}
  \delta_{f,k + p}.
\end {equation}

In the electron-LO phonon interaction one phonon of momentum $ {\bf q} $ is 
emitted, an electron $ k $ is annihilated, and an electron $ k' $ is created. 
The interaction of the electrons with the polarization created by the 
LO phonons is given by
\begin {equation}
  \label {int-LO}
  \int d {\bf r} \int dz \, \frac { 4 \pi e \gamma }{ | {\bf q} | }
  \hat {\Psi}^{e \dagger}({\bf r},z) \hat {B}({\bf r}, z) \hat \Psi^{\rm e}({\bf r},z),
\end {equation}
where $ \gamma $ is some interaction constant (see for example Ref. 
\cite {GaL}, \cite {Mah}), and the phonon field operator is
\begin {equation}
  \label {LO-fld}
  \hat {B}({\bf r}, z) =
  \sum_{{\bf q}}
  \left( \frac {1}{ 2 V \rho \omega_{o} } \right)^{1/2}
  e^{i {\bf q}_{\perp} \cdot {\bf r} + i q z} 
  \left( \hat {a}_{{\bf q}} + \hat {a}_{-{\bf q}}^{\dagger} \right).
\end {equation}
The components of $ {\bf q} $ are defined in the same manner as those of 
$ {\bf f} $, $ \hat {a}_{{\bf q}} $ is the annihilation operator, and $ \rho $ 
is the reduced mass per unit cell in the lattice.

The matrix element $ M^{\rm e-LO} $ can be found by sandwiching interaction
(\ref {int-LO}) between the states 
  $ \langle n(k)=0, n(k')=1, N(\omega_{o})=1 | $ 
and
  $ | n(k)=1, n(k')=0, N(\omega_{o})=0 \rangle $. Carrying out
the integration over the $ z $ coordinates, the square of the matrix element
is
\begin {equation}
  \label {|M-e-ph|^2}
  \left| M^{\rm e-LO}_{k,k',{\bf q}} \right|^{2} = \frac {1}{| {\bf q} |^{2}}
  \frac {4 \pi \alpha \omega_{o}^{3/2}}{V (2 m_{\rm e})^{1/2}}
  \left|
    \int d {\bf r} \left| \phi^{\rm e} ({\bf r}) \right|^{2}
       e^{-i {\bf q}_{\perp} \cdot {\bf r}}
  \right|^{2} 
  \delta_{k - k',q},
\end {equation}
where we exchanged $ \gamma $ for the known Fr\"olich constant $ \alpha $,
using the relation 
$ \gamma^{2} = (\rho \omega_{o}^{2}/ 4 \pi) 
               (2 \omega_{o} / m_{\rm e})^{1/2} (\alpha / e^{2}) $
(see Ref. \cite {Mah}).

Substituting $ \left| M^{\rm exc} \right|^{2} $ and 
$ \left| M^{\rm e-LO} \right|^{2} $ into equation (\ref {e-Be}), we can sum 
over $ p $ and $ {\bf q} $. Taking the normalizing volumes to infinity and thus 
exchanging the sums with integrals we obtain
\begin {eqnarray}
  \label {e-Be1}
  0 = \int \frac {d {\bf f}}{(2 \pi)^{3}}
  \left| M^{\rm exc}(\nu_{{\bf f}}) \right|^{2}
  2 \pi 
  \delta (\nu_{{\bf f}} - \epsilon_{k} - \varepsilon_{\rm g} - 
          \varepsilon_{f-k})
  (1 - n(k)) N({\bf f}) - \nonumber \\
  \int \frac {dk'}{2 \pi}
  \left| M^{\rm e-LO}(k-k') \right|^{2}
  2 \pi \delta (\epsilon_{k} - \epsilon_{k'} - \omega_{o})
  n (k),
\end {eqnarray}
where
\begin {equation}
  \label {|M-e-ep|^21}
  \left| M^{\rm exc}(\nu_{{\bf f}}) \right|^{2} = 
  \frac {2 \pi e^{2} p_{o}^{2}}{m_{o}^{2} \nu_{{\bf f}}}
  \left|
    \int d {\bf r} \, \phi^{\rm e}({\bf r}) \phi^{\rm h}({\bf r})
  \right|^{2},
\end {equation}
\begin {equation}
  \label {|M-e-ph|^21}
  \left| M^{\rm e-LO}(k-k') \right|^{2} =
  \frac {1}{2\pi} 
  \frac {4 \pi \alpha \omega_{o}^{3/2}}{(2 m_{\rm e})^{1/2}}
  \sigma(k-k'),
\end {equation}
and  
\begin {equation}
  \label {sigma}
  \sigma(k-k') =
  \int d {\bf r} \int d {\bf r}' 
  \left| 
    \phi^{\rm e}({\bf r}) \right|^{2} \left| \phi^{\rm e}({\bf r}') 
  \right|^{2}
  K_{0} \left( |k-k'||{\bf r} - {\bf r}'| \right).
\end {equation}
The zeroth order Bessel function $ K_{0} $ is a result of the integration over 
$ {\bf q}_{\perp} $. 

Since the matrix elements (\ref {|M-e-ep|^21}) and (\ref {|M-e-ph|^21}) are 
smooth functions of their arguments, and we are interested only in processes 
which involve electrons that were excited close to the threshold, we may take 
them at the threshold values. The threshold value of $ \nu_{{\bf f}} $ is 
$ \bar {\nu} = \varepsilon_{\rm g} + (1 + \eta) \omega_{o} $, where 
$ \eta = m_{\rm e}/m_{\rm h} $. This is the frequency that will excite 
electrons from the valence band to the threshold in the conduction band. 
The threshold value of $ |k-k'| $ is 
$ q_{o} = (2 m_{\rm e} \omega_{o})^{1/2} $. 

The electron wave functions that appear in $ \sigma $ limit the spatial 
integration to $ r \leq a $, where $ a $ is the wire width, therefore we can 
estimate that $ \sigma(q_{o}) \propto |\ln (q_{o} a)| $ for $ q_{o} a \ll 1 $ 
and $ \sigma(q_{o}) \propto 1/(q_{o} a)^{2} $ for $ q_{o} a \gg 1 $.
For GaAs $ q_{o} = 2.5 \times 10^{6} {\rm cm}^{-1} $, thus for most wire
widths we are in the regime of $ q_{o} a \gg 1 $, in which $ \sigma $
is expected to decrease with increasing wire cross section. The constant 
$ \alpha \sigma(q_{o}) \equiv \alpha^{*} $ is the effective Fr\"ohlich constant
in the one-dimensional case. According to Ref. \cite {KnR93} 
$ \alpha^{*} \approx 0.1 $ (which is of the same order as the bulk 
Fr\"ohlich constant) for a wire of cross section 
$ 200 \times 100 {\rm \AA}^{2} $, while according to Ref. \cite {RRL95} 
$ \alpha^{*} \approx 0.02 $ for a wire of cross section 
$ 300 \times 100 {\rm \AA}^{2} $.

We now perform the integrations which appear in equation (\ref {e-Be1}). We 
begin with the generating term $ {\cal G}(\epsilon_{k}) $, 
the free term (one that doesn't include $ n(k) $) in the balance equation 
(\ref {e-Be1}) for the electron occupation number. This term is physically the
generation rate of electrons in the conduction band. Since $ f \ll k $ we 
can approximate $ \varepsilon_{f-k} \approx \varepsilon_{-k} $. Using 
$ d {\bf f} = 
  \nu_{{\bf f}}^{2} \, d \nu_{{\bf f}} \, d \Omega_{{\bf f}}/c^{3} \approx
  \bar {\nu}^{2} \, d \nu_{{\bf f}} \, d \Omega_{{\bf f}}/c^{3} $, where 
$ d \Omega_{{\bf f}} $ stands for the solid angle increment, we obtain
\begin {eqnarray}
  \label {e-Be1-ep}
  \int \frac {d {\bf f}}{(2 \pi)^{3}}
  \left| M^{\rm exc}(\bar {\nu}) \right|^{2}
  2 \pi 
  \delta (\nu_{{\bf f}} - \epsilon_{k} - \varepsilon_{\rm g} - \varepsilon_{-k})
  N({\bf f}) = \nonumber \\
  2 \pi \left| M^{\rm exc}(\bar {\nu}) \right|^{2}
  D(\bar {\nu})
  \langle N(\epsilon_{k} + \varepsilon_{\rm g} + \varepsilon_{-k}) \rangle 
  \equiv {\cal G}(\epsilon_{k}),
\end {eqnarray}
where $ D(\bar {\nu}) = \bar {\nu}^{2} / (2 \pi^{2} c^{3}) $,
is the photon density of states, and
\begin {equation}
  \label {<>-def}
  \langle N(\epsilon_{k} + \varepsilon_{\rm g} + \varepsilon_{-k}) \rangle = 
  \int \left. \frac {d \Omega_{{\bf f}}}{4 \pi} N({\bf f}) 
  \right |_{\nu_{{\bf f}} = 
  \epsilon_{k} + \varepsilon_{\rm g} + \varepsilon_{-k}},
\end {equation}
is the angular average of $ N({\bf f}) $, for photons of frequency 
$ \epsilon_{k} + \varepsilon_{\rm g} + \varepsilon_{-k} $, that create 
electrons with energy $ \epsilon_{k} $.

We consider a narrow band photon excitation field such that the angular average
of the occupation number is given by
\begin {equation}
  \label {<N>}
  \langle N(\nu) \rangle = I \left( \frac {c}{\bar {\nu}} \right)^{3}
  \frac {\Delta \nu_{o}/2 \pi}{(\nu - \nu_{o})^{2} +
    \Delta \nu_{o}^{2}/4},          
\end {equation}
where $ I $ is the excitation field energy density, $ \Delta \nu_{o} $ is the 
spectral width, and $ \nu_{o} $ is the central frequency. We are interested in 
a narrow band excitation close to the threshold, therefore 
$ \Delta \nu_{o} \ll \omega_{o} $, and the detuning (the 
difference between the central excitation frequency and the threshold 
frequency) is small: 
$ \tilde {\nu}_{o} = \nu_{o} - \bar {\nu} \ll \omega_{o} $.

The integrations that appear in the expression for the decay rate 
$ \Gamma(\epsilon_{k}) $, the term that includes all coefficients of $ -n(k) $ 
in the kinetic equation (\ref {e-Be1}), will now be carried out. The first 
contribution to $ \Gamma(\epsilon_{k}) $ is that of the photons, and is 
equal to the generating term (\ref {e-Be1-ep}). The second contribution to 
$ \Gamma(\epsilon_{k}) $ is due to LO phonons,
\begin {equation}
  \label {e-Be1-ph}
  \Gamma_{\rm LO} (\epsilon_{k}) =
  \int \frac {dk'}{2 \pi} 
  \left| M^{\rm e-LO} (q_{o}) \right|^{2} 
  2 \pi \delta (\epsilon_{k} - \epsilon_{k'} - \omega_{o}) = 
  2 \alpha^{*} \omega_{o} 
  \left( \frac {\omega_{o}}{\epsilon_{k} - \omega_{o}} \right)^{1/2}.
\end {equation}

The photon contribution to the decay rate is negligible compared to the phonon 
contribution since radiative processes are slow compared to non-radiative 
processes. Therefore $ \Gamma(\epsilon_{k}) $ can be written as
\begin {equation}
  \label {decay-ek}
  \Gamma(\epsilon_{k}) = \Gamma_{\rm LO} (\epsilon_{k}) =
  \left( \frac {\Gamma_{c}^{3}}{\epsilon_{k} - \omega_{o}} \right)^{1/2},
\end {equation}
where
\begin {equation}
  \label {Gamma-c}
  \Gamma_{c} = (2 \alpha^{*})^{2/3} \omega_{o}.
\end {equation}
The meaning of the energy scale $ \Gamma_{c} $ will be explained at the end of 
this section. For $ \alpha^{*} = 0.02 $ \cite {RRL95} one obtains
$ \Gamma_{c} = 4 {\rm meV} $, thus 
$ \Gamma_{c} \ll \omega_{o} = 36 {\rm meV}$. 

Below the threshold, where $ \Gamma_{\rm LO} = 0 $, the acoustical phonon 
contribution to $ \Gamma $ is important. Since the latter is a smooth function 
near the threshold, a good approximation for $ \Gamma $ is 
$ \Gamma(\epsilon_{k}) = \tau_{\rm ac}^{-1} = const $ at 
$ \epsilon_{k} < \omega_{o} $. 

The occupation number of the electrons that were excited above the threshold is
given by expression (\ref {n-a}), where $ {\cal G} $ and $ \Gamma $ are given 
by expressions (\ref {e-Be1-ep}) and (\ref {decay-ek}) respectively.
It is clear from the expressions above that $ n(k) $ depends on $ k $ through
$ \epsilon_{k} $ only.

We now turn our attention to the kinetic equation for the luminescence photons,
\begin {eqnarray}
  \label {lp-Be}
  & &
  \frac {\partial}{\partial t} N(f,m,n) = 
  \sum_{k}
  \left| M^{\rm lum}_{k,f,m,n} \right|^{2}
  2 \pi
  \delta (\nu_{f,m,n} - \epsilon_{k} - \varepsilon) \times \nonumber \\ 
  & &
  \left[ 
    (1 + N(f,m,n)) n(\epsilon_{k}) \bar {n}(\varepsilon) -
    N(f,m,n) (1 - n(\epsilon_{k})) (1 - \bar {n}(\varepsilon))
  \right], 
\end {eqnarray}
which for $ \bar {n}(\varepsilon) = 1 $ and in case of spontaneous 
luminescence is
\begin {equation}
  \label {lp-Be-sp}
  \frac {\partial}{\partial t} N(f,m,n) = 
  \sum_{k}
  \left| M^{\rm lum}_{k,f,m,n} \right|^{2}
  2 \pi
  \delta (\nu_{f,m,n} - \epsilon_{k} - \varepsilon) 
  n(\epsilon_{k}).
\end {equation}
$ M^{\rm lum} $ is the matrix element for the luminescence
photon-electron interaction. The time derivative in the equation above is kept 
in order to clarify what is the luminescence source.

Let us first look at the matrix element that describes the recombination of an
electron with a hole from a specific impurity. We will then have to sum over 
all impurities and average over all impurity configurations. In this process a 
photon of quantum numbers $ f,m,n $ is emitted, and an electron $ k $ and a 
hole in impurity $ i $ are destroyed. The interaction is given by
\begin {equation}
  \label {int-lum}
  \int d {\bf r} \int dz \, \frac {e p_{o}}{m_{o} c} 
  \hat \Psi^{\rm e}({\bf r}, z) \hat {A}^{\rm lum}({\bf r}, z) 
  \hat \Psi^{\rm h}({\bf r}, z),
\end {equation}
where 
\begin {equation}
  \label {lum-fld}
  \hat {A}^{\rm lum}({\bf r}, z) =
  \sum_{f,m,n} 
  \left( \frac {2 \pi c^{2}}{\nu_{f,m,n}} \right)^{1/2}
  \left( \chi_{f,m,n} ({\bf r},z) \hat {a}_{f,m,n} +
         \chi_{f,m,n}^{*} ({\bf r},z) \hat {a}_{f,m,n}^{\dagger} 
  \right)
\end {equation}
is the field operator of the luminescence photons.

The matrix element of the interaction above between the states
$\langle N(f,m,n)=1, n(\epsilon_{k})=0, \bar {n}(\varepsilon)=0 |$ and
$| N(f,m,n)=0, n(\epsilon_{k})=1, \bar {n}(\varepsilon)=1
  \rangle$ for holes in impurity $ i $ is
\begin {eqnarray}
  \label {M-lp-e}
  M^{\rm lum}_{k,f,m,n}(i) = \frac {e p_{o}}{m_{o} c} 
  \left( \frac {2 \pi c^{2}}
               {\pi R^{2} L^{2} \nu_{f,m,n} 
                  \left| J_{m+1}(\kappa_{m,n} R) \right|^{2} }
  \right)^{1/2}
  \times \nonumber \\
  \int d {\bf r} \int dz \,
  e^{i k z} \phi^{\rm e}({\bf r}) \psi ({\bf r} - {\bf r}_{i}, z - z_{i}) 
  e^{- i f z - i m \varphi} J^{*}_{m}(\kappa_{m,n} r).
\end {eqnarray}

We now sum the matrix element squared over all impurities and average over 
all impurity configurations. This yields
\begin {eqnarray}
  \label {|M-lp-e|^2}
  & &
  \left| M^{\rm lum}_{k,f,m,n} \right|^{2} = 
  \left( \frac {e p_{o}}{m_{o} c} \right)^{2}
  \frac {2 \pi c^{2}}
        {\pi R^{2} L^{2} \nu_{f,m,n} 
          \left| J_{m+1}(\kappa_{m,n} R) \right|^{2} }
  \times \nonumber \\
  & &
  \sum_{i} \frac {1}{V} \int d {\bf r}_{i} \int dz_{i}
  \left|
    \int d {\bf r} \int dz \,  
    e^{i k z} \phi^{\rm e}({\bf r})
    \psi ({\bf r} - {\bf r}_{i}, z - z_{i})
    e^{- i f z - i m \varphi} 
    J^{*}_{m}(\kappa_{m,n} r) \right|^{2}.
\end {eqnarray}
Carrying out the integrations over the $ z $ coordinates we obtain 
\begin {eqnarray}
  \label {|M-lp-e|^21}
  \left| M^{\rm lum}_{k,f,m,n} \right|^{2} = 
  \left( \frac {e p_{o}}{m_{o} c} \right)^{2}
  \frac {2 \pi c^{2}}
        {\pi R^{2} L^{2} \nu_{f,m,n} 
          \left| J_{m+1}(\kappa_{m,n} R) \right|^{2} } L
  \times \nonumber \\
  \sum_{i} \frac {1}{V} \int d {\bf r}_{i}
  \left|
    \int d {\bf r} \,  
    \phi^{\rm e}({\bf r})
    \psi ({\bf r} - {\bf r}_{i}, -k)
    e^{- i m \varphi} 
    J^{*}_{m}(\kappa_{m,n} r) \right|^{2},
\end {eqnarray}
where $ \psi ({\bf r}, k) = \int dz \, e^{- i k z} \psi ({\bf r}, z) $. 
The second argument of $ \psi $, $ f - k $ was replaced by $ -k $ because
$ f \ll k $.

Since the impurities are randomly distributed in the bulk, the average over
all impurity configurations of the product of the hole wave functions, that
appear in (\ref {|M-lp-e|^21}), can depend only on the difference between 
their coordinates. Thus we define
\[
  {\cal L}({\bf r} - {\bf r}', -k) \equiv 
  \int d {\bf r}_{i} \,
  \psi ({\bf r} - {\bf r}_{i}, -k) \psi^{*} ({\bf r}' - {\bf r}_{i}, k).
\]
The summation over impurities divided by the normalizing volume renders a
factor of $ n_{\rm imp} $ - the impurity spatial density. The averaged matrix 
element squared can then be written as
\begin {eqnarray}
  \label {|M-lp-e|^22}
  \left| M^{\rm lum}_{k,f,m,n} \right|^{2} = 
  \left( \frac {e p_{o}}{m_{o} c} \right)^{2}
  \frac {2 \pi c^{2}}
        {\pi R^{2} L \nu_{f,m,n} 
          \left| J_{m+1}(\kappa_{m,n} R) \right|^{2} } \,
  n_{\rm imp}
  \int d {\bf r} \int d {\bf r}' \times 
  \nonumber \\
  \phi^{\rm e}({\bf r}) \phi^{{\rm e} *}({\bf r}') {\cal L}({\bf r} - {\bf r}', -k)
  e^{- i m (\varphi-\varphi')} 
  J^{*}_{m}(\kappa_{m,n} r) J_{m}(\kappa_{m,n} r').
\end {eqnarray}

The matrix element squared is much larger for $ m = 0 $ than for other values 
of $ m $, since $ r $ and $ r' $ are constrained to a small region in the 
$ xy $ plane (due to the electron wave functions), and $ J_{0} $ is the only 
Bessel function which is finite at $ r \rightarrow 0 $, 
$ \left. J_{0}(r) \right|_{r \rightarrow 0} = 1 $. Thus the matrix element
squared is proportional to $ \delta_{m,0} $. From now on the index $ m $ is
omitted with the understanding that we are dealing only with $ m = 0 $.

The matrix element squared is a smooth function of $ \nu_{f,n} $ and $ k $,
therefore these can be substituted by their threshold values. The threshold 
values of $ \nu_{f,n} $ and $ k $ are 
$ \bar {\nu}' = \varepsilon + \omega_{o} $ and
$ k_{o} = (2 m_{\rm e} \omega_{o})^{1/2} $ respectively. The kinetic equation 
for the luminescence photons can then be written as
\begin {equation}
  \label {lp-Be1}
  \frac {\partial}{\partial t} N(f,n) = 
  \int \frac {dk}{2 \pi}
  \left| M^{\rm lum}_{n}(k_{o}) \right|^{2}
  2 \pi
  \delta (\nu_{f,n} - \epsilon_{k} - \varepsilon) 
  n(\epsilon_{k}), 
\end {equation} 
where the matrix element squared is
\begin {eqnarray}
  \label {|M-lp-e|^23}
  \left| M^{\rm lum}_{n}(k_{o}) \right|^{2} =
  \left( \frac {e p_{o}}{m_{o} c} \right)^{2}
  \frac {2 \pi c^{2}}
        {\pi R^{2} L \bar {\nu}
           \left| J_{1}(\kappa_{n} R) \right|^{2} } \,
  n_{\rm imp} L
  \times \nonumber \\
  \int d {\bf r} \int d {\bf r}' \,
  \phi^{\rm e}({\bf r}) \phi^{{\rm e} *}({\bf r}') {\cal L}({\bf r} - {\bf r}', - k_{o}).
\end {eqnarray}

Substituting $ n(\epsilon_{k}) $ into equation (\ref {lp-Be1}) and integrating 
over $ k $ we obtain
\begin {equation}
  \label {lp-source}
  \frac {\partial}{\partial t} N(f,n) =   
  \left| M^{\rm lum}_{n}(k_{o}) \right|^{2}
  \left( \frac {2 m_{\rm e}}{\nu_{f,n} - \varepsilon} \right)^{1/2}
  \frac {{\cal G}(\nu_{f,n} - \varepsilon)}
        {\Gamma(\nu_{f,n} - \varepsilon)}. 
\end {equation}

The luminescence source can be characterized by the spectral dependence of 
$ E(\nu') d \nu' $, the energy of the emitted field, of all spectral modes 
within the interval $ d \nu' $, per unit length of the wire, per unit time. 
Multiplying the generating term of the luminescence photons (the right hand 
side of equation (\ref {lp-source})) by 
$ \delta (\nu' - \nu_{f,n}) \nu' d \nu' / L $ and summing over $ f $ and $ n $,
we obtain
\begin {equation}
  E(\nu') \, d \nu' = \frac {\nu'}{L} \sum_{n} \sum_{f}
  \left| M^{\rm lum}_{n}(k_{o}) \right|^{2}
  \left( \frac {2 m_{\rm e}}{\nu_{f,n} - \varepsilon} \right)^{1/2}
  \frac {{\cal G}(\nu_{f,n} - \varepsilon)}
        {\Gamma(\nu_{f,n} - \varepsilon)} \,
  \delta (\nu' - \nu_{f,n}) \, d \nu'.
\end {equation}

The variable $ \nu_{f,n} $ in the square root can be replaced with its
threshold value $ \varepsilon + \omega_{o} $, due to the smoothness of this
function. Taking the normalization volume (that appears in the matrix element)
to infinity the sums over $ f $ and $ n $ are transformed into integrals. 
Performing the integration we obtain the final result
\begin {equation}
  \label {E-Be}
  E(\nu') = {\cal C} \, \nu' D(\nu') \xi (\nu'),
\end {equation}
where 
\begin {equation}
  \label {C}
  {\cal C} = \left( \frac {e p_{o}}{m_{o} c} \right)^{2}
             \frac {2 \pi c^{2}}{\bar {\nu}'} n_{\rm imp}
             \int d {\bf r} \int d {\bf r}' \, 
             \phi^{\rm e}({\bf r}) \phi^{{\rm e} *}({\bf r}')
             {\cal L} ({\bf r} - {\bf r}', -k_{o}),
\end {equation}
is a constant, and
\begin {equation}
  \label {xi-Be}
  \xi(\nu') = \left( \frac {2 m_{\rm e}}{\omega_{o}} \right)^{1/2}
  \frac {{\cal G}(\nu' - \varepsilon)}
        {\Gamma(\nu' - \varepsilon)} =
  \left( \frac {2 m_{\rm e}}{\omega_{o}} \right)^{1/2}
  \frac {2 \pi \left| M^{\rm exc}(\bar {\nu}) \right|^{2} 
        D(\bar {\nu})}
        {\Gamma(\nu' - \varepsilon)}
  \langle N((1 + \eta)(\nu' - \varepsilon) + \varepsilon_{\rm g})\rangle,
\end {equation}
is essentially the product of the electron generation rate and the electron 
life-time at the corresponding energy.

\begin{figure}
  \begin{center}
    \setlength{\unitlength}{0.27in}
    \begin{picture}(10,7)(0,0)
      \put(0,0){\psfig {figure=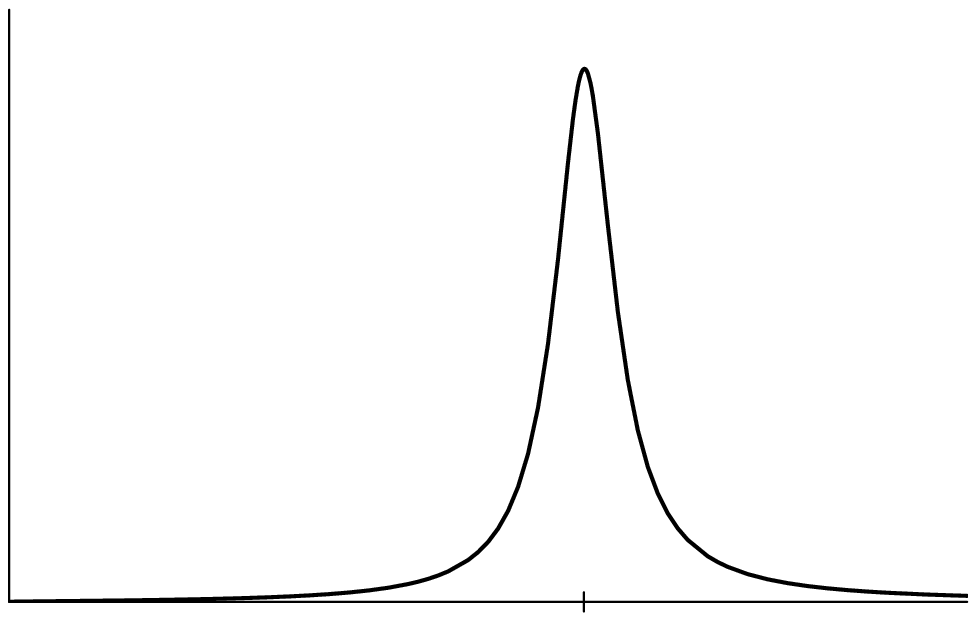,width=10\unitlength}}
      \put(6,-0.1){\makebox(0,0)[t]{$\tilde{\nu}'_{o}$}}
      \put(0,6.5){\makebox(0,0)[b]{$\xi(\nu')$}}
      \put(10.3,0){\makebox(0,0)[lb]{$\nu' - \bar {\nu}'$}}
      \put(0,0){\makebox(0,0)[tr]{$0$}}
      \put(5.6,2.75){\makebox(0,0)[r]{$\rightarrow$}}
      \put(6.4,2.75){\makebox(0,0)[l]{$\leftarrow$}}
      \put(7.2,2.75){\makebox(0,0)[l]{$\Delta \nu_{o}'$}}
      \put(7.3,4.7){\psfig {figure=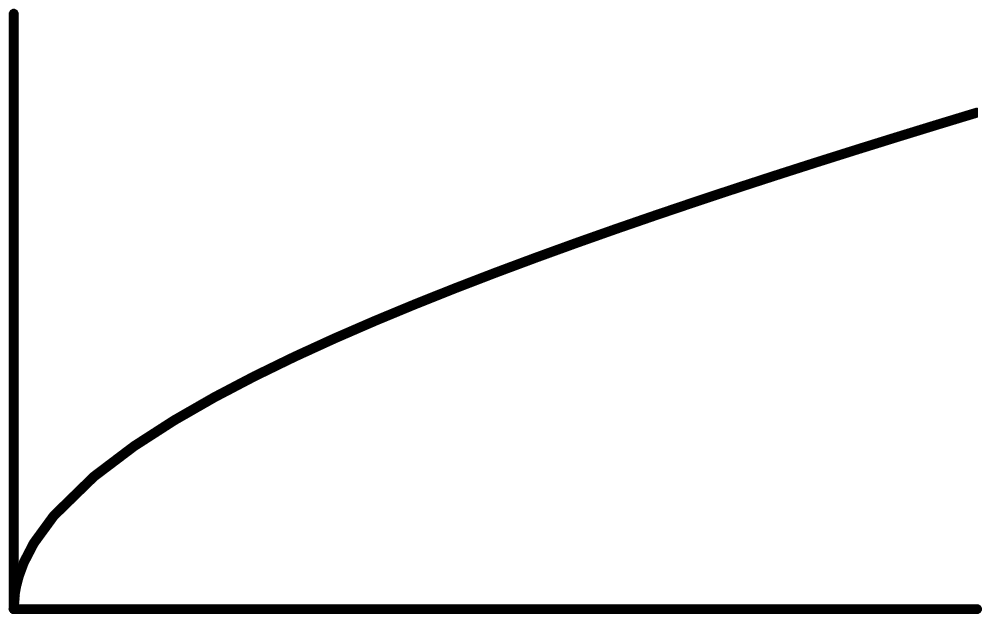,width=2.5\unitlength}}
      \put(7.3,6.5){\makebox(0,0)[b]{$\xi(\nu')$}}
      \put(10,4.7){\makebox(0,0)[lb]{$\nu' - \bar {\nu}'$}}
      \put(7.3,4.7){\makebox(0,0)[tr]{$0$}}
    \end{picture}
  \end{center}
  \caption {The behavior of $ \xi(\nu') $ as a function of $ \nu' $ for
            $ \tilde {\nu}_{o} \gg \Delta \nu_{o} $ ($ \eta = 1/3 $ and
            $ \tilde {\nu}_{o} = 8 \Delta \nu_{o} $), where
            $ \tilde {\nu}'_{o} = \tilde {\nu}_{o} / (1 + \eta) $ and
            $ \Delta \nu'_{o} = \Delta \nu_{o} / (1 + \eta) $.
            An enlargement of its behavior
            for $ \nu' - \bar {\nu}' \ll \tilde {\nu}'_{o} $ is
            shown in the inset.}
  \label {xi1}
\end {figure}

Since $ \nu' D(\nu') $ is a smooth function of $ \nu' $, the spectral 
distribution of the luminescence is given by $ \xi(\nu') $. Let us first 
consider a narrow band excitation that does not overlap the threshold, i.e. 
$ \tilde {\nu}_{o} \gg \Delta \nu_{o} $. In this case $ \xi(\nu') $ is a peaked 
function of $ \nu' $, reproducing the shape of the excitation. It is of a
Lorentzian shape of width 
$ \Delta \nu'_{o} = \Delta \nu_{o} / (1 + \eta) $, centered at 
$ \bar {\nu}' + \tilde {\nu}'_{o} $, where 
$ \tilde {\nu}_{o}' = \tilde {\nu}_{o} / (1 + \eta) $ (see Fig. \ref {xi1})
\begin {equation}
  \label {xi-Lor}
  \xi (\nu') \propto 
  \left( \frac {\tilde {\nu}_{o}}{(1 + \eta)^{3} \Gamma_{c}^{3}} \right)^{1/2}
  \frac { \Delta \nu'_{o} / (2 \pi)}
        {\left( \nu' - \bar {\nu}' - \tilde {\nu}_{o}' \right)^{2}
          + \left( \Delta \nu_{o}'/ 2 \right)^{2}}.
\end {equation}
The rescaling of $ \nu' $ compared to $ \nu $ follows from the obvious
relations (see Fig. \ref {qwr}) 
$ \epsilon_{k} + \varepsilon_{k} =\nu - \varepsilon_{\rm g} $ and
$ \nu' = \varepsilon + \epsilon_{k} $.

Only close to the threshold, when 
$ \nu' - \bar {\nu}' \ll \tilde {\nu}_{o} / (1 + \eta) $, the spectral
behavior of the luminescence differs from that of the excitation and is given 
by (see the inset of Fig. \ref {xi1})
\begin {equation}
 \label {sqrt}
 \xi(\nu') \propto 
 \frac {\Delta \nu_{o}}{2 \pi \tilde {\nu}_{o}^{2} \Gamma_{c}^{3/2}} \,
                   (\nu' - \bar {\nu}')^{1/2}.
\end {equation}

\begin {figure}
  \begin{center}
    \setlength{\unitlength}{0.69in}
    \begin{picture}(4,3)(0,0)
      \put(0,0){\psfig {figure=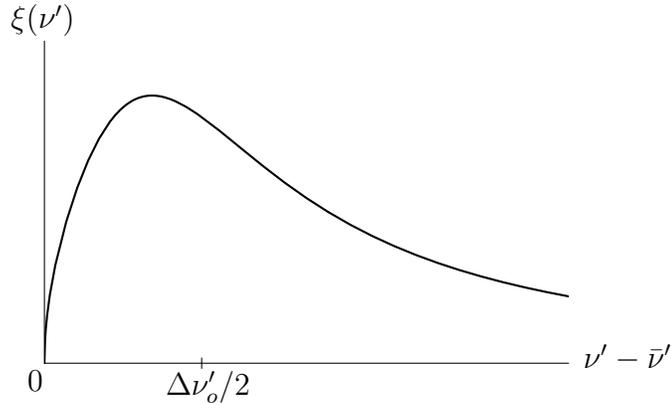,width=4\unitlength}}
      \put(1.25,0){\makebox(0,0)[t]{$\Delta \nu_{o}'/2$}}
      \put(0,2.6){\makebox(0,0)[b]{$\xi(\nu')$}}
      \put(4.1,0.1){\makebox(0,0)[l]{$\nu' - \bar {\nu}'$}}
      \put(0,0){\makebox(0,0)[tr]{$0$}}
    \end{picture}
  \end{center}
  \caption {The behavior of $ \xi(\nu') $ as a function of $ \nu' $ for
            $ \Delta \nu_{o} \gg \tilde {\nu}_{o} $ ($ \eta = 1/3 $ and 
            $ \Delta \nu_{o} = 8 \tilde {\nu}_{o}$), where 
            $ \Delta \nu'_{o} = \Delta \nu_{o} / (1 + \eta) $.}
  \label {xi3}
\end {figure}

When the narrow band excitation overlaps the threshold, i.e.\ 
$ \tilde {\nu}_{o} \ll \Delta \nu_{o} $, the spectral distribution of the luminescence
is very different from that of the excitation (see Fig. \ref {xi3}):
\begin {equation}
  \xi(\nu') \propto \frac {2}{\pi \Delta \nu_{o} \, \Gamma_{c}^{3/2}} \,
  (\nu' - \bar {\nu}')^{1/2}
  \;\;\;\;\;
  \mbox {when} 
  \;\;\;
  \nu' - \bar {\nu}' \ll \frac {\Delta \nu_{o}'}{2},
\end {equation}
and
\begin {equation}
  \xi (\nu') \propto 
  \frac {\Delta \nu_{o}}{2 \pi (1 + \eta)^{2} \Gamma_{c}^{3/2}} \, 
                     (\nu - \bar {\nu}')^{-3/2}
  \;\;\;\;\;
  \mbox {when} 
  \;\;\; 
  \nu' - \bar {\nu}' \gg \frac {\Delta \nu_{o}'}{2}.
\end {equation}

If one is interested in luminescence due to electrons that were excited
close to but {\em below} the threshold, via an excitation that is centered 
{\em above} the
threshold, i.e.\ $ \nu' - \bar {\nu}' < 0 $ and $ \tilde {\nu}_{o} > 0 $, then
$ \Gamma(\nu' - \varepsilon) $ in expression (\ref {xi-Be}) should be replaced
by $ \tau_{\rm ac}^{-1} $. In this case $ \xi(\nu') $ has the same Lorentzian
form as (\ref {xi-Lor}) with the amplitude
$ \left( \tilde {\nu}_{o} / ((1 + \eta)^{3} \Gamma_{c}^{3}) \right)^{1/2} $
replaced by $ \tau_{\rm ac} / (1 + \eta) $. 

One can expect that the luminescence distribution $ \xi(\nu') $, derived from
the BE, is not correct for frequencies $ \nu' $ close to
the threshold $ \bar {\nu}' $. Luminescence of such frequencies is due to
electrons with energies $ \epsilon_{k} $ near the threshold, that have a large 
width $ \Gamma(\epsilon_{k} $), (see equation (\ref {decay-ek})), while the 
BE assumes that the width of the states is small compared to 
other energy scales of the problem. In our case one of these energy scales is 
$ \epsilon_{k} - \omega_{o} $. Equating the width with this energy scale,
$ \epsilon_{k} - \omega_{o} = \Gamma(\epsilon_{k}) $, one finds an additional
energy scale of the problem (and a critical decay rate), that is given by
$ \Gamma_{c} $ that has been defined before (\ref {Gamma-c}).

When $ \epsilon_{k} - \omega_{o} \lesssim \Gamma_{c} $, we have  
$ \Gamma(\epsilon_{k}) \gtrsim \epsilon_{k} - \omega_{o} $, and the assumption 
of the BE breaks down. One would therefore expect that for 
$ \epsilon_{k} - \omega_{o} \lesssim \Gamma_{c} $ in the description of the 
excited electrons, and for $ \nu' - \bar {\nu}' \lesssim \Gamma_{c} $ in the 
description of the hot luminescence, the results predicted by the Boltzmann 
equation would fail. We will learn from the QKEs that in 
fact the situation is more complicated.

\subsection {The Quantum Description}
\label {Q-Des}
As we have seen in the previous section, the Boltzmann description may fail for
electrons excited to the vicinity of the threshold and hence for luminescence 
photons emitted by these electrons. Therefore, close to the threshold, the 
electrons and the luminescence photons have to be treated using QKEs.
These QKEs will be written by applying the generalization rules to equations 
(\ref {e-Be1}) and (\ref {lp-Be1}). It is assumed that the energies of the 
holes, the phonons and the excitation photons are not broadened.

In the generalization to the QKEs only those terms that were kept in the 
Boltzmann description will be retained. Threshold values of the arguments of 
the matrix elements will be substituted as was done in the Boltzmann 
description, due to the smoothness of the matrix elements.

The QKE for the electrons is
\begin {eqnarray}
  \label {e-qke}
  0 & = & \int \frac {d {\bf f}}{(2 \pi)^{3}}
  \left| M^{\rm exc}(\bar {\nu}) \right|^{2}
  2 \pi 
  \delta (\nu_{{\bf f}} - \epsilon - \varepsilon_{\rm g} - \varepsilon_{-k})
  N({\bf f}) - \nonumber \\
  & & 
  \int \frac {dk'}{2 \pi} \int d \epsilon' \,
  \delta (\Gamma(k', \epsilon')| \epsilon' - \epsilon_{k'} -
                                 \Delta \epsilon (k', \epsilon'))
  \left| M^{\rm e-LO}(q_{o}) \right|^{2}
  2 \pi \delta (\epsilon - \epsilon' - \omega_{o})
  n (k, \epsilon),
\end {eqnarray}
where $ \epsilon $ is the electron off-shell energy, $ n $ is the electron 
occupation function, and $ \Gamma $ and $ \Delta \epsilon $ are the electron 
energy level width and shift, respectively.

The coefficients of $ - n(k, \epsilon) $ in equation (\ref {e-qke}) can be 
recognized as the electron level width due to LO phonons
\begin {eqnarray}
  \label {Gamma-e}
  \Gamma_{\rm LO} (k, \epsilon) = 
  \int \frac {dk'}{2 \pi} \int d \epsilon' \,
  \delta (\Gamma(k', \epsilon')| \epsilon' - \epsilon_{k'} -
                                 \Delta \epsilon(k', \epsilon'))
  \left| M^{\rm e-LO}(q_{o}) \right|^{2}
  2 \pi \delta (\epsilon - \epsilon' - \omega_{o}),
\end {eqnarray}
and the total electron width is given by
\begin {equation}
  \label {Gamma-tot}
  \Gamma(k, \epsilon) = \Gamma_{\rm LO} + 1 / \tau_{\rm ac}.
\end {equation}
The photon contribution to the level width was neglected (as it was neglected 
in the Boltzmann description of the decay term).

One can see that the generating term in equation (\ref {e-qke}) is equal to 
expression (\ref {e-Be1-ep}), the generating term of the electrons' BE, 
with the on-shell energy $ \epsilon_{k} $ replaced by the off-shell energy 
$ \epsilon $ (in the exciting photon occupation number)
\begin {equation}
  \label {generating}
  {\cal G}(k, \epsilon) = 
  2 \pi \left| M^{\rm exc}(\bar {\nu}) \right|^{2}
  D(\bar {\nu})
  \langle N(\epsilon + \varepsilon_{\rm g} + \varepsilon_{-k}) \rangle.
\end {equation}

The occupation function $ n(k,\epsilon) $ is found by substituting 
$ {\cal G} $ and $ \Gamma $ into (\ref {n-ae}). The generating term 
$ {\cal G}(k, \epsilon) $ depends on $ k $ through 
$ \varepsilon_{-k} = \eta \epsilon_{k} $ only, therefore 
$ {\cal G}(k, \epsilon) = {\cal G}(\epsilon_{k}, \epsilon) $. It follows from 
(\ref {Gamma-e}) that $ \Gamma $ is independent of $ k $ (and so is the level 
shift). Therefore, $ n (k, \epsilon) = n(\epsilon_{k}, \epsilon) $.

\begin {figure}
  \begin {center}
    \setlength{\unitlength}{0.1165in}
    \begin{picture}(22,15)
      \put(0,0){\psfig {figure=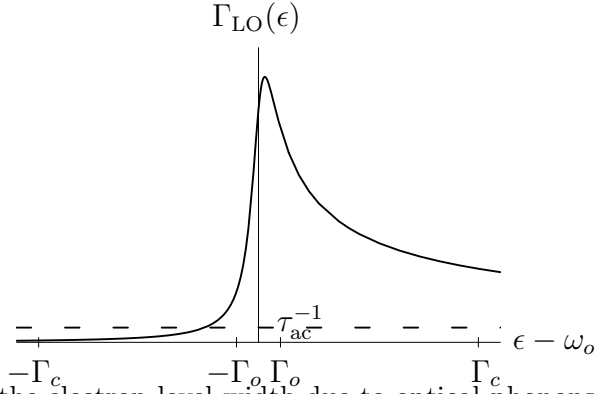,width=22\unitlength}}
      \put(1,-0.5){\makebox(0,0)[t]{$-\Gamma_{c}$}}
      \put(10,-0.5){\makebox(0,0)[t]{$-\Gamma_{o}$}}
      \put(12.3,-0.5){\makebox(0,0)[t]{$\Gamma_{o}$}}
      \put(21.3,-0.5){\makebox(0,0)[t]{$\Gamma_{c}$}}
      \put(11,14.5){\makebox(0,0)[b]{$\Gamma_{\rm {LO}}(\epsilon)$}}
      \put(22.5,0){\makebox(0,0)[lb]{$\epsilon - \omega_{o}$}}
      \put(11.9,1.2){\makebox(0,0)[l]{$\tau_{\rm ac}^{-1}$}}
    \end{picture}
  \end{center}
  \caption {The behavior of the electron level width due to optical phonons,
            at the vicinity of the threshold for LO phonon emission, when the
            electron level width at the bottom of the subband is considered.}
  \label {Gamma}
\end {figure}

Equation (\ref {Gamma-e}) is a self consistent equation for the level width.
Integrating the right hand side of the equation we find that
\begin {eqnarray}
  \label {Gamma-te}
  \Gamma_{\rm LO} (\epsilon) = 
  \alpha^{*} \omega_{o}^{3/2}
  \left\{ \left[ \epsilon - \omega_{o} - 
          \Delta \epsilon (\epsilon - \omega_{o}) + 
          \frac {i}{2} 
          \Gamma (\epsilon - \omega_{o}) \right]^{-1/2} + 
  \right. \nonumber \\
  \left. \left[ \epsilon - \omega_{o} -
         \Delta \epsilon (\epsilon - \omega_{o}) - 
         \frac {i}{2} 
         \Gamma (\epsilon - \omega_{o}) \right]^{-1/2} 
  \right\}.
\end {eqnarray}
When $ \epsilon - \omega_{o} \ll \omega_{o} $ the arguments of 
$ \Gamma (\epsilon - \omega_{o}) $ and
$ \Delta \epsilon (\epsilon - \omega_{o}) $ are close to the bottom of the 
band. In this case the level shift can be neglected since it is just a
small renormalization. The level width close to the bottom of the band
$ \Gamma_{o} $, is due to scattering with acoustic phonons and due to thermal
recombination. The latter contribution to $ \Gamma_{o} $ can be neglected,
while the acoustic phonon contribution can be considered as a constant (though 
not the same as the width due to acoustic phonons at the threshold). We assume 
that $ \Gamma_{o} $ is small compared to all other energy scales of the 
problem. Thus close to the threshold
\begin {equation}
  \label {Gamma-th}
  \Gamma_{\rm LO} (\epsilon) = 
  \alpha^{*} \omega_{o}^{3/2}
  \left\{ \left[ \epsilon - \omega_{o} + 
          \frac {i}{2} \Gamma_{o} \right]^{-1/2} +
          \left[ \epsilon - \omega_{o} -
          \frac {i}{2} 
          \Gamma_{o} \right]^{-1/2} 
  \right\}.
\end {equation}
The behavior of $ \Gamma_{\rm LO} (\epsilon) $ when $ \epsilon $ is 
close to $ \omega_{o} $ is shown in Fig. \ref {Gamma}.

From expression (\ref {Gamma-th}) we see that above the threshold for 
$ \epsilon - \omega_{o} \gg \Gamma_{o}/2 $, $ \Gamma (\epsilon) $ 
can be written as
\begin {equation}
  \label {Gamma-<}
  \Gamma (\epsilon) = \Gamma_{\rm LO}(\epsilon) = 
  2 \alpha^{*} \omega_{o}
  \left( \frac {\omega_{o}}{\epsilon - \omega_{o}} \right)^{1/2} =
  \left ( \frac {\Gamma_{c}^{3}}{\epsilon - \omega_{o}} \right)^{1/2}.
\end {equation}
Comparing expression (\ref {Gamma-<}) to expression (\ref {decay-ek}), it is
evident that above the threshold $ \Gamma (\epsilon) $ is equal to 
$ \Gamma(\epsilon_{k}) $ when the on-shell energy $ \epsilon_{k} $ is replaced 
by the off-shell energy $ \epsilon $. Since $ \Gamma_{o} $ is the smallest 
energy scale in the problem, expression (\ref {Gamma-<}) can be used for the 
electron level width above the threshold.

Contrary to the Boltzmann description, the electron-LO phonon scattering
contributes to the electron level width below the threshold as well.
When $ \omega_{o} - \epsilon \gg \Gamma_{o} $ the electron width due to 
optical phonons decays like $ |\epsilon - \omega_{o}|^{-3/2} $, much
faster than the decay above the threshold (see Fig. \ref {Gamma}).
The contribution of the acoustic phonons to the electron width below the 
threshold $ \tau_{\rm ac}^{-1} $, will be negligible compared to 
$ \Gamma_{\rm LO} $ for 
$ \omega_{o} - \epsilon \ll (\Gamma_{o} \tau_{\rm ac})^{2/3} \Gamma_{c} / 2 $.
Since $ \Gamma_{o} \tau_{\rm ac} \ll 1 $ and $ \Gamma_{c} \tau_{\rm ac} \gg 1 $
the right hand side of the inequality above is much smaller than $ \Gamma_{c} $,
and much larger than $ \Gamma_{o} $ (see the crossing point of 
$ \Gamma_{\rm LO}(\epsilon) $ and $ \tau_{\rm ac}^{-1} $ in Fig. \ref {Gamma}).

The QKE that is obtained by applying the 
generalization rules to the BE for the luminescence photons (\ref {lp-Be1}) is
\begin {eqnarray}
  \label {lp-qke}
  & &
  \frac {\partial}{\partial t} 
  \left\{ \left( \frac {1}{2} + N(f,n,\nu') \right) 
          \delta (\gamma(f,n,\nu') | \nu' - \nu_{f,n} - 
                                           \Delta \nu(f,n,\nu')) 
  \right\} = \nonumber \\
  & & 
  \delta (\gamma(f,n,\nu') | \nu' - \nu_{f,n} - 
                                   \Delta \nu(f,n,\nu')) \times
  \nonumber \\
  & &
  \int \frac {dk}{2 \pi} 
  \int d \epsilon 
  \left| M^{\rm lum}_{n}(k_{o}) \right|^{2}
  \delta (\Gamma(\epsilon) | \epsilon - \epsilon_{k} -
                             \Delta \epsilon(\epsilon))
  2 \pi \delta (\nu' - \epsilon - \varepsilon) 
  n(\epsilon_{k}, \epsilon), 
\end {eqnarray}
where $ \nu' $ is the photon off-shell energy, $ N $ is the photon 
occupation function, and $ \gamma $ and $ \Delta \nu $ are the photon energy 
level width and shift, respectively. All the photon functions are written for 
$ m = 0 $.

By applying the generalization rules to the right hand side of the full 
BE (\ref {lp-Be}) one can find the photon level width
\begin {eqnarray}
  \label {gamma}
  \gamma(n,\nu') = 
  - \int \frac {dk}{2 \pi} \int d\epsilon
  \left| M^{\rm lum}_{n}(k_{o}) \right|^{2}
  \delta (\Gamma (\epsilon)| \epsilon - \epsilon_{k} -
  \Delta \epsilon (\epsilon)) \times \nonumber \\
  2 \pi \delta (\nu' - \epsilon - \varepsilon)
  n(\epsilon_{k},\epsilon).  
\end {eqnarray}
Note that $ \gamma < 0 $ corresponds to photon generation.
It is evident from expression (\ref {gamma}) that $ \gamma $ does not depend
on $ f $, and therefore neither does $ \Delta \nu $.

Using (\ref {gamma}) equation (\ref {lp-qke}) can be written as
\begin {eqnarray}
  \label {lp-qke1}
  \frac {\partial}{\partial t} 
  \left\{ \left( \frac {1}{2} + N(f,n,\nu') \right) 
          \delta (\gamma(n,\nu') | \nu' - \nu_{f,n} - 
                                           \Delta \nu(n,\nu')) 
  \right\} = \nonumber \\
  - \gamma(n,\nu')
  \delta (\gamma(n,\nu') | \nu' - \nu_{f,n} - 
                                           \Delta \nu(n,\nu')).
\end {eqnarray}
The right hand side of equation (\ref {lp-qke1}) is the photon generating term,
the term from which the luminescence spectral distribution will be obtained.

As in the classical case, the luminescence source can be characterized by 
the spectral dependence of $ E(\nu') d \nu'$, that is obtained by multiplying 
the right hand side of equation (\ref {lp-qke1}) by $ \nu' d \nu' / L $ and 
summing over $ f $ and $ n $
\begin {equation}
  \label {E-nu}
  E (\nu') \, d \nu' = - \frac {\nu'}{L} \sum_{n} \sum_{f} \gamma(n,\nu')
  \delta (\gamma (n,\nu') | \nu' - \nu_{f,n} - \Delta \nu (n,\nu')) \, d \nu'.
\end {equation}
The main contribution to $ E(\nu') $ comes from $ f $ and $ n $ such that 
$ \nu_{f,n} $ is close to $ \nu' $.

It is convenient to write $ \gamma(n,\nu') $ in the following form
\begin {equation}
  \label {gamma-1}
  \gamma (n,\nu') = - \frac {1}
                           {\pi R^{2} \left| J_{1}(\kappa_{n} R) \right|^{2}}
  \, {\cal C} \, \xi (\nu'),
\end {equation}
where $ {\cal C} $ is given by (\ref {C}), and
\begin {equation}
  \label {bar-gam}
  \xi(\nu') = \int \frac {dk}{2 \pi} \int d \epsilon \,
  \delta (\Gamma (\epsilon)| \epsilon - \epsilon_{k} -
  \Delta \epsilon (\epsilon))
  2 \pi \delta (\nu' - \epsilon - \varepsilon)
  n(\epsilon_{k},\epsilon).
\end {equation}
When the electrons' energy is taken to be on shell, that is the electron 
smeared delta function is reduced to a singular delta function, 
$ \xi(\nu') $ is reduced to expression (\ref {xi-Be}). In writing $ \gamma $
in the form (\ref {gamma-1}), the normalization volume has been separated from 
the rest of the function, that is independent of the index $ n $, and can 
therefore be taken out of the summation in $ E (\nu') $.

Taking the normalization volume to infinity the sums over $ f $ and $ n $ are
transformed into integrals. From equation (\ref {gamma-1}) it is evident 
that $ \left. \gamma \right|_{R \rightarrow \infty} \rightarrow 0 $.
This is due to the fact that the photons are emitted into an "infinite" space,
while their interaction with the electrons is confined to the finite volume
of the wire. In this case $ \gamma $ and $ \Delta \nu $ are negligible compared
to the other widths, and the smeared photon delta function that appears in the 
expression for $ E (\nu') $ becomes a singular delta function, giving
\begin {equation}
  \label {E-nu1}
  E (\nu') = {\cal C} \, \nu' \xi(\nu')
  \int_{-\infty}^{\infty} \frac {df}{2 \pi} 
  \int_{0}^{\infty} \frac {\kappa \, d \kappa}{2 \pi} \,
  \delta (\nu' - \nu_{f,n}) = 
  {\cal C} \, \nu' D(\nu') \xi(\nu').
\end {equation}
Since $ \nu' D (\nu') $ is a smooth function it is clear at this point that in 
order to understand the dependence of $ E (\nu') $ on $ \nu' $, the behavior of 
$ \xi (\nu') $ should be analyzed.

Performing the integration over $ \epsilon $ in (\ref {bar-gam}), and 
substituting the expression we got for $ n(\epsilon_{k}, \epsilon) $, we get
\begin {eqnarray}
  \label {bar-gam1}
  \xi(\nu') = (2 m_{\rm e})^{1/2}
  \frac 
    {2 \pi \left| M^{\rm exc}(\bar {\nu}) \right|^{2} D (\bar {\nu})}
    {\Gamma (\nu' - \varepsilon)} 
  \int_{0}^{\infty} 
  \frac {d \epsilon_{k}}{\epsilon_{k}^{1/2}} \,
  \delta (\Gamma(\nu' - \varepsilon)| \nu' - \varepsilon - \epsilon_{k})
  \times \nonumber \\
  \langle N(\nu' - \varepsilon + \varepsilon_{k} + \varepsilon_{\rm g}) 
  \rangle.
\end {eqnarray}
Since $ \epsilon_{k} - \omega_{o} \ll \omega_{o} $, we substitute 
$ \omega_{o} $ for $ \epsilon_{k} $ in the one-dimensional density of states. 
This integral contains in fact two Lorentzians. The first is of width 
$ \Gamma(\nu'-\varepsilon) $: The electron level width at energy 
$ \epsilon = \nu' - \varepsilon $. The second Lorentzian is of width 
$ \Delta \nu_{o}/ \eta $: A width that is proportional to the spectral width 
of the exciting photon source. Since the excitation is such that 
$ \epsilon_{k} $ is close to $ \omega_{o} $, and all the energy scales that can 
characterize the level width are much smaller than $ \omega_{o} $, the 
integrand goes to zero when $ \epsilon_{k} $ approaches zero from above. Thus, 
the lower boundary of the integral can be taken to minus infinity. As a result 
the spectral distribution of the luminescence is given by the function
\begin {eqnarray}
  \label {bar-gam2}
  & &
  \xi(\nu') = \left( \frac {2 m_{\rm e}}{\omega_{o}} \right)^{1/2}
  \frac {2 \pi \left| M^{\rm exc}(\bar {\nu}) \right|^{2} 
         D (\bar {\nu})}
        {\Gamma(\nu' - \varepsilon)}
  I \left( \frac {c}{\bar {\nu}} \right)^{3}
  \times \nonumber \\
  & &
  \int_{-\infty}^{\infty} 
  d \epsilon_{k}
  \left( \frac {\Gamma(\nu' - \varepsilon)/2 \pi}
               {(\nu' - \varepsilon - \epsilon_{k})^{2} + 
         \Gamma(\nu' - \varepsilon)^{2}/4} \right)
  \left( \frac {\Delta \nu_{o}/ 2 \pi}
        {[( \nu' - \varepsilon + \eta \epsilon_{k} + \varepsilon_{\rm g} ) -
            \nu_{o}]^{2} + \Delta \nu_{o}^{2}/4} \right).
\end {eqnarray}

The region of validity of the BE is evident from equation
(\ref {bar-gam2}). In order to obtain the result of the BE
(\ref {xi-Be}), one has to replace the first Lorentzian with a delta function.
This can be done only when its width is smaller than that of the second
Lorentzian, i.e. $ \Gamma(\nu'- \varepsilon) \ll \Delta \nu_{o} / \eta $. It 
follows from this inequality that the Boltzmann description of the luminescence
spectra is correct only far from the threshold, when 
$ \nu' - \bar {\nu}' \gg \eta^{2} \Gamma_{c}^{3}/\Delta \nu_{o}^{2} 
  \equiv \nu_{c} $.
The quantum interval $ \nu_{c} $ differs from the naive estimate $ \Gamma_{c} $
(see the discussion at the end of subsection \ref {B-Des}): It is 
larger for "narrow" band excitation ($ \Delta \nu_{o} \ll \Gamma_{c} $), and 
smaller for "wide" band excitation ($ \Delta \nu_{o} \gg \Gamma_{c} $).

For an excitation that is mostly of frequencies above 
$ \bar {\nu} + \nu_{c} $, the spectral dependence of the energy of the
emitted photons will behave according to the predictions of the BE
(see Figs. \ref {xi1}-\ref {xi3}). However, as long as the 
excitation is above the threshold, very close to the threshold $ \xi(\nu') $ 
will increase linearly with $ \nu' $, and not as a square root, see 
(\ref {sqrt}).

Non-Boltzmann behavior is obtained when 
$ \Gamma(\nu' - \varepsilon) \gg \Delta \nu_{o} / \eta $, that is
$ \nu' - \bar {\nu}' \ll \nu_{c} $, and the second Lorentzian can be treated 
as a delta function. In this case one obtains
\begin {equation}
  \label {xi-q}
  \xi(\nu') \propto
  \frac {\eta}{2 \pi (1 + \eta)^{2}}
  \left[
     \left( \nu' - \bar {\nu}' - \frac {\tilde {\nu}_{o}}{1 + \eta} \right)^{2}
     + \frac {\eta^{2}}{4 (1 + \eta)^{2}} \,
       \frac {\Gamma_{c}^{3}}{\nu' - \bar {\nu}'} \right]^{-1}.
\end {equation}
We bring here two specific examples of extreme non-Boltzmann behavior of 
$ \xi(\nu') $. In both examples the excitation is centered within the quantum 
interval, i.e. $ \tilde {\nu}_{o} \ll \nu_{c} $, and "narrow" band, i.e. 
$ \Delta \nu_{o} \ll \Gamma_{c} $. Due to the latter inequality 
$ \nu_{c} \gg \Gamma_{c} $.

\begin {figure}
  \begin{center}
    \setlength{\unitlength}{0.15in}
    \begin{picture}(20,10)(0,0)
      \put(0,0){\psfig {figure=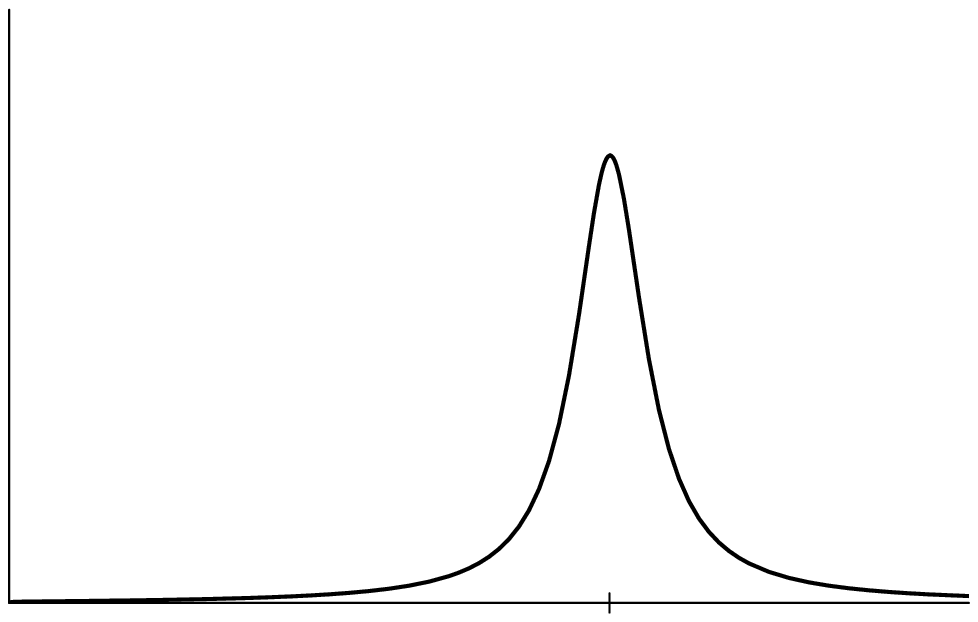,width=18\unitlength}}
      \put(0,11.5){\makebox(0,0)[b]{$\xi(\nu')$}}
      \put(18.5,0){\makebox(0,0)[lb]{$\nu' - \bar {\nu}'$}}
      \put(11.3,-0.1){\makebox(0,0)[t]{$\tilde {\nu}_{o}'$}}
      \put(0,0){\makebox(0,0)[tr]{$0$}}
      \put(10.5,4.3){\makebox(0,0)[r]{$\rightarrow$}}
      \put(12,4.3){\makebox(0,0)[l]{$\leftarrow$}}
      \put(13.5,4.3){\makebox(0,0)[l]{$\Delta \nu_{o}'$}}
      \put(13,8){\psfig {figure=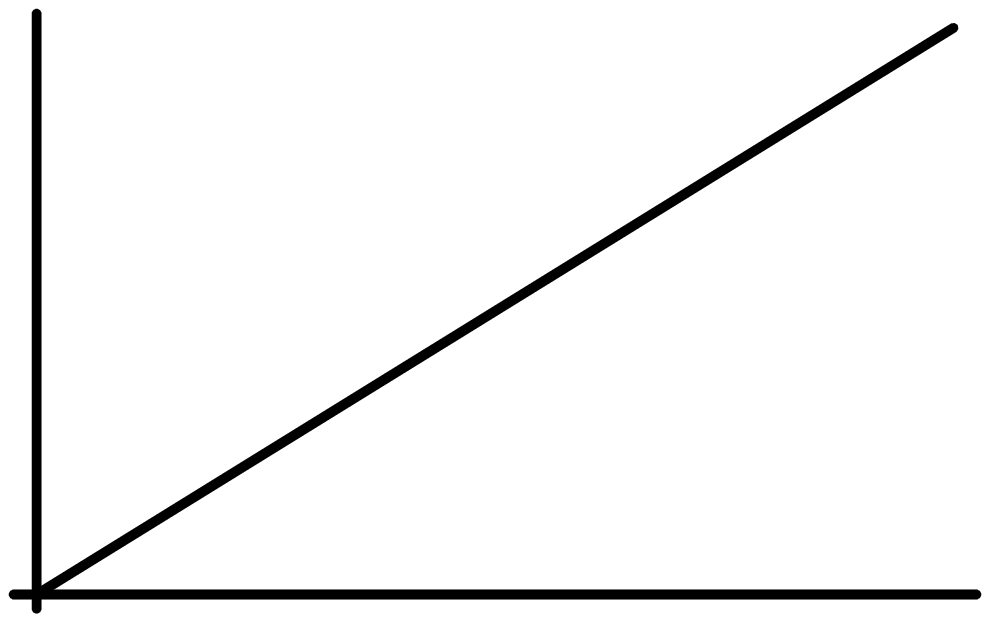,width=5\unitlength}}
      \put(13,11.5){\makebox(0,0){$\xi(\nu')$}}
      \put(13,8){\makebox(0,0)[tr]{$0$}}
      \put(18.5,8){\makebox(0,0)[lb]{$\nu' - \bar {\nu}'$}}	
   \end{picture}
  \end{center}
  \caption {The behavior of $ \xi(\nu') $ as a function of $ \nu' $ for
            $ \nu_{c} \gg \tilde {\nu}_{o} \gg \Gamma_{c} \gg \Delta \nu_{o} $
            ($ \eta = 1/3, \; \tilde {\nu}_{o} = 15 \Delta \nu_{o}, $ and 
            $ \Gamma_{c} = 5 \Delta \nu_{o} $), where 
            $ \tilde {\nu}'_{o} = \tilde {\nu}_{o} / (1 + \eta) $ and
            $ \Delta \nu'_{o} = 
              \eta \,
              (\Gamma_{c}^{3} / \tilde {\nu}'_{o} )^{1/2} 
              / (1 + \eta)  $. 
            An enlargement of its behavior for 
            $ \nu' - \bar {\nu}' \ll 
              (\eta^{2}/4) \Gamma_{c}^{3}/\tilde {\nu}_{o}^{2} $ 
            is shown in the inset.}
  \label {xi-q1}
\end {figure}

In the first case the detuning is large, $ \tilde {\nu}_{o} \gg \Gamma_{c} $, therefore
the energy width of an electron excited by the central excitation frequency,
$ \epsilon_{k} = \omega_{o} + \tilde {\nu}_{o} / (1 + \eta) $, that is of the order of
$ ( \Gamma_{c}^{3} / \tilde {\nu}_{o} )^{1/2} $, is small compared to its
distance from the threshold. This is true for most of the electrons. In such a
case the electron states are "well defined", but the prediction of the 
BE for the electron distribution is wrong, since its width 
$ \Delta \nu_{o} / (1 + \eta) $ is much smaller than the width of the states.
As a result it follows from (\ref {xi-q}) that the luminescence spectra is 
symmetric and centered at the classical position 
$ \bar {\nu}' + \tilde {\nu}'_{o} $, with
$ \tilde {\nu}'_{o} = \tilde {\nu}_{o} / (1+ \eta) $, but its width is given by 
the quantum-mechanical width of the electron states (see Fig. \ref {xi-q1})
\[
  \Delta \nu'_{o} = \frac {\eta}{1 + \eta}
                    \left[
                      \frac {\Gamma_{c}^{3}}{\tilde {\nu}_{o} / (1 + \eta)}
                    \right]^{1/2}.
\]

\begin {figure}
  \begin{center}
    \setlength{\unitlength}{0.1in}
    \begin{picture}(30,16)(0,0)
      \put(0,0){\psfig {figure=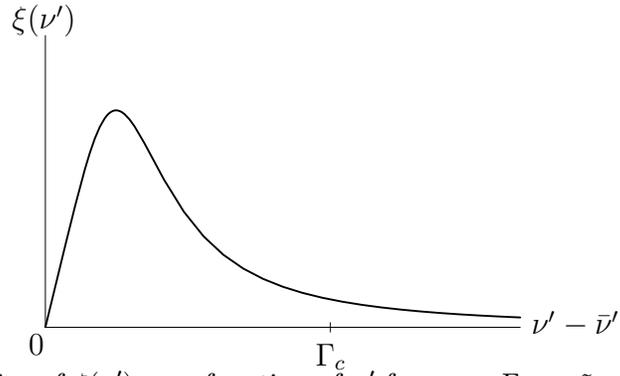,width=25\unitlength}}
      \put(0,16){\makebox(0,0)[b]{$\xi(\nu')$}}
      \put(25.5,0){\makebox(0,0)[lb]{$\nu' - \bar {\nu}'$}}
      \put(15,0){\makebox(0,-0.5)[t]{$\Gamma_{c}$}}
      \put(0,0){\makebox(0,0)[tr]{$0$}}
   \end{picture}
  \end{center}
  \caption {The behavior of $ \xi(\nu') $ as a function of $ \nu' $ for
            $ \nu_{c} \gg \Gamma_{c} \gg \tilde {\nu}_{o}, \Delta \nu_{o} $
            ($ \eta = 1/3, \; \tilde {\nu}_{o} = 2 \Delta \nu_{o}, $ and 
            $ \Gamma_{c} = 15 \Delta \nu_{o} $).}
  \label {xi-q3}
\end {figure}

In the second case the detuning is small, $ \tilde {\nu}_{o} \ll \Gamma_{c} $, 
and the electrons are excited to "badly defined" states, since the broadening 
of these states is larger than their distance from the threshold. As a result 
the luminescence spectral distribution differs greatly from that predicted by 
the BE. One finds from (\ref {xi-q}) that (see Fig. \ref {xi-q3})
\[
  \xi(\nu') \propto \frac {2}{\pi \eta \Gamma_{c}^{3}} \, (\nu' - \bar {\nu}')
  \;\;\;\;\;
  \mbox {when} 
  \;\;\;
  \nu' - \bar {\nu}' \ll \Gamma_{c},
\]
and
\[
  \xi(\nu') \propto \frac {\eta}{2 \pi (1 + \eta)^{2}} \,
                    \frac {1}{(\nu' - \bar {\nu}')^{2}}
  \;\;\;\;\;
  \mbox {when} 
  \;\;\; 
  \nu_{c} \gg \nu' - \bar {\nu}'\gg \Gamma_{c}.
\]

In conclusion let us note that the hole dispersion plays an essential role in 
determining the quantum behavior of the luminescence spectra. The width of the 
second Lorentzian in (\ref{bar-gam2}) $ \Delta \nu_{o}/ \eta $, is not equal to
the width of the classical luminescence line $ \Delta \nu_{o}/(1+\eta) $,
and the integral is not a simple convolution of the classical luminescence 
profile with the spectral function of an electron state.
One can see from (\ref {bar-gam2}) that for a flat hole band ($\eta =0$) the 
second Lorenzian does not depend on the integration variable and the first 
Lorenzian is integrated to one, restoring the BE result, unexpectedly.

The luminescence due to electrons that were excited {\em below} the threshold
($ \epsilon < \omega_{o} $) by an excitation centered {\em above} the threshold
($ \tilde {\nu}_{o} > 0 $) is not obtained by the simple exchange of 
$ \Gamma_{\rm LO} $ by $ \tau_{\rm ac}^{-1} $, as in the classical case.
As long as 
$ \nu' - \bar {\nu}' < (\Gamma_{o} \tau_{\rm ac})^{2/3} \Gamma_{c} / 2 $
one should take $ \Gamma = \Gamma_{\rm LO} $ from expression 
(\ref {Gamma-th}) since 
$ \Gamma_{\rm LO} (\nu' - \varepsilon) > \tau_{\rm ac}^{-1} $, and therefore
although the electrons are below the threshold, it is the optical phonon
contribution to the electron width that is dominant. 

\section {Conclusions}
We have presented a recipe that allows one to generalize the BE to a QKE. 
The QKE obtained by employing this recipe is the same as that obtained by
the Keldysh Green function technique in the self-consistent Born approximation.
The advantage of this method for writing the QKE is that it provides a physical
understanding of the terms in the QKE, and it allows one to neglect those terms
that were negligible in the Boltzmann description. This is due to the fact that
the equations are written for quantities that are similar to the quantities
described by the Boltzmann equation.

We considered the specific example of hot luminescence from a QWR. The recipe 
described above was used in order to generalize the set of BEs that describe
the problem to a set of QKEs. Solving these equations we were able to describe
the luminescence spectral distribution.

We have shown that there is a domain of luminescence frequencies, that 
correspond to a domain of phtoexcited electron energies, for which the quantum 
description of the luminescence spectral distribution leads to a different 
behavior than that given by the Boltzmann description. This quantum domain 
could not be easily guessed from level width considerations. Two other 
nontrivial conclusions were obtained. The first is the role played by the hole 
mass in the definition of the quantum domain. When the hole dispersion relation
is flat the quantum domain shrinks to zero and the results of the Boltzmann 
description are retrieved. The second nontrivial conclusion is that there is 
an energy domain below the threshold for LO phonon emission, in which the LO 
phonon contribution to the electron level width is dominant.

\acknowledgements
One of the authors (Y.L.) is indebted to P. W\"olfle for drawing his attention
to the importance of QKEs in optical phenomena. The 
research was supported by the Israel Academy of Sciences.


\end {document}